\documentclass{vldb}
\usepackage{hadianeh}
\usepackage{graphicx}
\usepackage{balance} 
\usepackage{macros}
\usepackage{comment}
\usepackage{algorithm2e}
\usepackage{wrapfig}

\usepackage{cite}
\usepackage{algpseudocode}
\usepackage{balance}

\usepackage[font={small,sf,bf}]{caption}
\usepackage{hyperref}

\vldbTitle{In-RDBMS Hardware Acceleration of Advanced Analytics}
\vldbAuthors{Divya Mahajan, Joon Kyung Kim, Jacob Sacks, Adel Ardalan, Arun Kumar, and Hadi Esmaeilzadeh}
\vldbDOI{https://doi.org/10.14778/3236187.3236188}
\vldbVolume{11}
\vldbNumber{11}
\vldbYear{2018}

\begin{document}

% ****************** TITLE ****************************************

\title{In-RDBMS Hardware Acceleration of Advanced Analytics}

% possible, but not really needed or used for PVLDB:
%\subtitle{[Extended Abstract]
%\titlenote{A full version of this paper is available as\textit{Author's Guide to Preparing ACM SIG Proceedings Using \LaTeX$2_\epsilon$\ and BibTeX} at \texttt{www.acm.org/eaddress.htm}}}

% ****************** AUTHORS **************************************


\numberofauthors{6} %  in this sample file, there are a *total*

\author{Divya Mahajan$^\ast$ \quad Joon Kyung Kim$^\ast$ \quad Jacob Sacks$^\ast$ \quad\quad Adel Ardalan$^\dag$\vspace{5pt} \\ 
Arun Kumar$^\ddag$ \quad Hadi Esmaeilzadeh$^\ddag$\vspace{5pt}\\
\affaddr{\normalsize $^\ast$Georgia Institute of Technology \quad\quad\quad $^\dag$University of Wisconsin-Madison \quad\quad\quad $^\ddag$University of California, San Diego}\vspace{3pt}\\
\url{{divya_mahajan,jkkim,jsacks}@gatech.edu}{} \quad\quad\quad \url{adel@cs.wisc.edu}{} \quad\quad\quad \url{{arunkk,hadi}@eng.ucsd.edu}{}
%
%\email{\{divya\_mahajan, jkkim, jsacks6\}@gatech.edu}
}

\maketitle

\begin{abstract}
The data revolution is fueled by advances in machine learning, databases, and hardware design.
Programmable accelerators are making their way into each of these areas independently.
As such, there is a void of solutions that enables hardware acceleration at the intersection of these disjoint fields.
%
%Although timely, there is a void of solutions that brings these disjoint directions together.
%
%Although timely, innovation in these areas is being explored independently, hence, there is a void of solutions that unifies these disjoint directions.
%
This paper sets out to be the initial step towards a unifying solution for in-\textbf{\sffamily D}atabase \textbf{\sffamily A}cceleration of Adva\textbf{\sffamily n}ced \textbf{\sffamily A}nalytics (\dana). 
%
%\dana empowers database users to leap beyond traditional data summarization techniques and seamlessly utilize hardware-accelerated machine learning.
%
Deploying specialized hardware, such as FPGAs, for in-database analytics currently requires hand-designing the hardware and manually routing the data.
Instead, \dana automatically maps a high-level specification of advanced analytics queries to an FPGA accelerator.
The accelerator implementation is generated for a User Defined Function (UDF), expressed as a part of an SQL query using a Python-embedded Domain-Specific Language (DSL). 
To realize an efficient in-database integration, \dana accelerators contain a novel hardware structure, \striders, that directly interface with the buffer pool of the database.
%
%process the data structures that result from query planning and execution. To do so, the accelerators  
%
%\dana obtains the schema and page layout information from the database catalog to configure the \striders.
%
\striders extract, cleanse, and process the training data tuples that are consumed by a multi-threaded FPGA engine that executes the analytics algorithm.
We integrate \dana with \psql to generate hardware accelerators for a range of real-world and synthetic datasets running diverse ML algorithms.
Results show that \dana-enhanced \psql provides, on average, \avgspeeduphcreal end-to-end speedup for real datasets, with a maximum of \maxspeeduphcreal. 
Moreover, \dana-enhanced \psql is, on average, \greenavgspeeduphcreal faster than the multi-threaded Apache MADLib running on Greenplum.
\dana provides these benefits while hiding the complexity of hardware design from data scientists and allowing them to express the algorithm in $\approx$30-60 lines of Python.

\end{abstract}

\section{Introduction}
\label{sec:intro}

\begin{figure}
	\centering
	\includegraphics[height=0.6\linewidth]{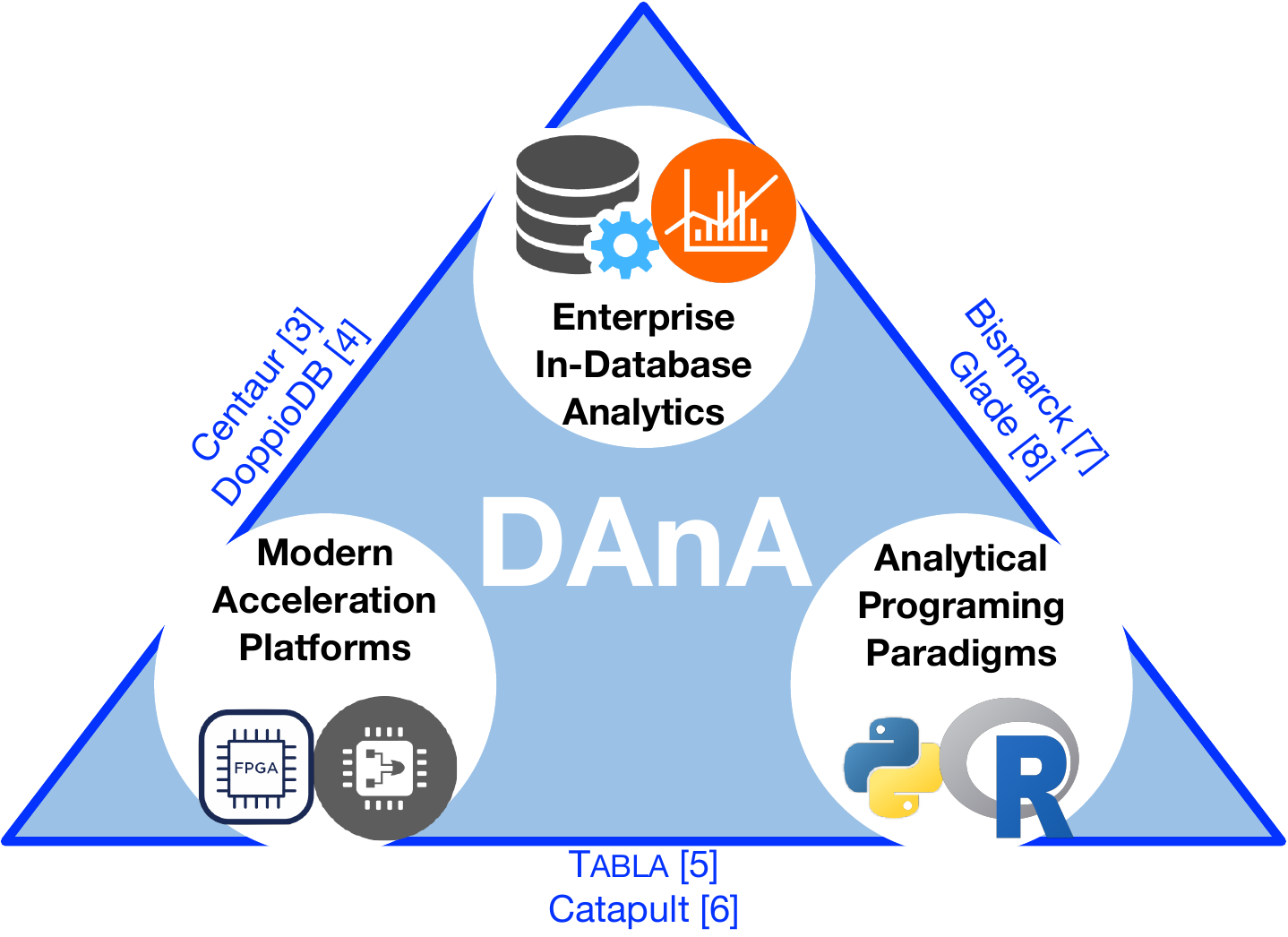}
	\vspace{-7pt}
	\caption{\dana represents the fusion of three research directions, in contrast with prior works~\cite{centaur, doppiodb, tabla:hpca, chimps, bismarck, glade} that merge two of the areas.
}
	\label{fig:triad}
	\vspace{-3.5ex}
\end{figure}

Relational Database Management Systems (RDBMSs) are the cornerstone of large-scale data management in almost all major enterprise settings. 
However, data-driven applications in such environments are increasingly migrating from simple SQL queries towards advanced analytics, especially machine learning (ML), over large datasets~\cite{gartner, sas}. 
As illustrated in Figure~\ref{fig:triad}, there are three concurrent and important, but hitherto disconnected, trends in this data systems landscape: (1) enterprise in-database analytics~\cite{centaur,doppiodb} , (2) modern hardware acceleration platforms~\cite{tabla:hpca, catapult}, and (3) programming paradigms which facilitate the use of analytics~\cite{bismarck, glade}.

The database industry is investing in the integration of ML algorithms within RDBMSs, both on-premise and cloud-based~\citecolored{aws_psql, azure_db}.
This integration enables enterprises to exploit ML without sacrificing the auxiliary benefits of an RDBMS, such as transparent scalability, access control, security, and integration with their business intelligence interfaces~\cite{odm, ore, madlib1, madlib2, msdm, bismarck, hogwild, glade, orion}.
Concurrently, the computer architecture community is extensively studying the integration of specialized hardware accelerators within the traditional compute stack for ML applications~\cite{tabla:hpca, chimps, cosmic:micro, dadiannao, pudiannao}. 
Recent work at the intersection of databases and computer architecture has led to a growing interest in hardware acceleration for relational queries as well.
This includes exploiting GPUs~\cite{hippogriffdb} and reconfigurable hardware, such as Field Programmable Gate Arrays (FPGAs)~\cite{centaur, doppiodb, gustavofpga, fpgapartitioning, gustavoce}, for relational operations.
Furthermore, cloud service providers like Amazon AWS~\cite{amazon-ec2-f1}, Microsoft Azure~\cite{haas}, and Google Cloud~\cite{tpu}, are also offering high-performance specialized platforms due to the potential gains from modern hardware.
Finally, the applicability and practicality of both in-database analytics and hardware acceleration hinge upon exposing a high-level interface to the user.
%
%Several research projects have investigated ways of simplifying in-database analytics~\cite{bismarck, hogwild, madlib2, glade, orion} and hardware acceleration~\cite{tabla:hpca} for machine learning by focusing on a particular subset of algorithms which use vanilla or stochastic gradient descent (SGD).
%
This triad of research areas are currently studied in isolation and are evolving independently.
Little work has explored the impact of moving analytics within databases on the design, implementation, and integration of hardware accelerators.
Unification of these research directions can help mitigate the inefficiencies and reduced productivity of data scientists who can benefit from  in-database hardware acceleration for  analytics. 
Consider the following example.

\begin{example}
\label{ex:ad}
\vspace{-3ex}
A marketing firm uses the Amazon Web Services (AWS) Relational Data Service (RDS) to maintain a \psql\ database of its customers.
A data scientist in that company forecasts the hourly ad serving load by running a multi-regression model across a hundred features available in their data. %that contains hundreds of features for ML and hundreds of GBs of training data.
%
%She decides to try various ML algorithms for regression, e.g., linear regression with feature interactions~\cite{hastie}, and artificial neural networks. 
%
Due to large training times, she decides to accelerate her workload using FPGAs on Amazon EC2 F1 instances~\cite{amazon-ec2-f1}.
%
%\footnote{\url{https://goo.gl/ooUD68}}.% instead of paying for far more instances to distribute her training.
%
%Currently, this is highly non-trivial, requiring her to know the principles and techniques of hardware design.
%
Currently, this requires her to learn a hardware description language, such as Verilog or VHDL, program the FPGAs, and go through the painful process of hardware design, testing, and deployment, individually for each ML algorithm.
%
%Since no major in-RDBMS vendor or cloud providers support hardware acceleration frameworks, she would have to manually extract, copy, and reformat her large dataset in order to use such research ML tools that enable FPGA acceleration~\cite{tabla, dnnweaver:micro, cosmic:micro}.
%
Recent research has developed tools to simplify FPGA acceleration for ML algorithms~\cite{tabla, dnnweaver:micro, cosmic:micro}.
%
%, which do not interface with or provide in-RDBMS support. 
%
%However, since no major in-RDBMS vendor or cloud providers support hardware acceleration, 
%
However, these solutions do not interface with or support RDBMSs, requiring her to manually extract, copy, and reformat her large dataset.
\vspace{-6pt}
\end{example}

To overcome the aforementioned roadblocks, we devise \dana, a \textit{cohesive stack that enables deep integration between FPGA acceleration and in-RDBMS execution of advanced analytics}.
\dana exposes a high-level programming interface for data scientists/analysts based on conventional languages, such as SQL and Python.
Building such a system requires: (1) providing an intuitive programming abstraction to express the combination of ML algorithm and required data schemas; and (2) designing a hardware mechanism that transparently connects the FPGA accelerator to the database engine for direct access to the training data pages.% while training the model.

To address the first challenge, \dana enables the user to express RDBMS User-Defined Functions (UDFs) using familiar practices of Python and SQL. 
The user provides their ML algorithm as an update rule using a Python-embedded Domain Specific Language (DSL), while an SQL query specifies data management and retrieval.
To convert this high level ML specification into an accelerated execution without manual intervention, we develop a comprehensive  stack. 
Thus, \dana is a solution that breaks the algorithm-data pair into software execution on the RDBMS for data retrieval and hardware acceleration for running the analytics algorithm.

With respect to the second challenge, \dana offers \striders, which avoid the inefficiencies of conventional Von-Neumann CPUs for data handoff by seamlessly connecting the RDBMS and FPGA.
\striders directly feed the data to the analytics accelerator by walking Jacthe RDBMS buffer pool. 
Circumventing the CPU alleviates the cost of data transfer through the traditional memory subsystem.
These \striders are backed with an Instruction Set Architecture (ISA) to ensure programmability and ability to cater to the variations in the database page organization and tuple length across different algorithms and training datasets.
%
%Thus, that can program the \strider to cater for the variability in database page layout. 
%
They are designed to ensure \emph{multi-threaded acceleration} of the learning algorithm to amortize the cost of data accesses across  concurrent threads.
\dana automatically generates the architecture of these accelerator threads, called execution engines, that selectively combine a Multi-Instruction Multi-Data (MIMD) execution model with Single-Instruction Multi-Data (SIMD) semantics to reduce the instruction footprint.
While generating this MIMD-SIMD accelerator, \dana tailors its architecture to the ML algorithm's computation patterns, RDBMS page format, and available FPGA resources.
As such, this paper makes the following technical contributions:

% and realizes a new dimension in the design of DNN accelerators
%
%To this end, \dana provides means for data scientists to exploit FPGAs for accelerating in-RDBMS advanced analytics without requiring expertise in hardware design or manual intervention in data management and makes the following technical contributions:

%
\begin{itempacked}

\item Merges three disjoint research areas to enable transparent and efficient hardware acceleration for in-RDBMS analytics. 
Data scientists with no hardware design expertise can use \dana to harness hardware acceleration without manual data retrieval and extraction whilst retaining familiar programming environments.

\item Exposes a high-level programming interface, which combines SQL UDFs with a Python DSL, to jointly specify training data and computation. 
This unified abstraction is backed by an extensive compilation workflow that automatically transforms the specification to an accelerated execution. % without programmer involvement in hardware design or data handling.   
%

%\item With \dana, database users can use the power of extensively established RDBMSs to efficiently collect, process, and analyze their large influx of data, whilst  still being exposed to the traditional interface of in-database analytics for learning, whereas the underlying analytics computation can be automatically accelerated using high-performance low-power FPGA architectures. 

\item Integrates an FPGA and an RDBMS engine through \striders that are a novel on-chip interfaces.
\striders bypass CPU to directly access the training data from the buffer pool, transfer this data onto the FPGA, and unpack the feature vectors and labels. %, as well as ML computations on the accelerator. 

\item Offers a novel execution model that fuses thread-level and data-level parallelism to execute the learning algorithm computations. 
%
%This model works symbiotically with \striders to exploit unique opportunities for dynamically interleaving data conversion and ML computation. 
%
This model exposes a domain specific instruction set architecture that offers automation while providing efficiency.

%\dana's hardware architecture contains an \emph{execution engine} that accelerates the ML computations and works symbiotically with \striders to exploit unique opportunities for dynamically interleaving data conversion and ML computation.
%
%Combined with \striders, this design enables \dana to exploit unique opportunities to dynamically interleave data conversion and ML computation.
%
%Both \striders and the execution engine are backed by their unique Instruction Set Architectures (ISA) that enables efficient execution and low-level programmability of the architecture.
%
%The execution engines supports a variable length selective Single Instruction Multiple Data (SIMD) ISA which aims to reduce the instruction footprint on the accelerator to divert the memory resources to store the training data.
%
\end{itempacked}

%\dana's malleable hardware architecture can be synthesized on the desired FPGA according to the ML algorithm, RDBMS page properties, and FPGA resources.
%
%To exploit this high-performance architecture, \dana offers a simple high-level DSL within Python for the users to  express a wide variety of popular ML algorithms. 
%
%\dana's domain-specific compiler and \hardwaregen~ then use this high-level specification and 
%
We prototype \dana with \psql to automatically accelerate the execution of several popular ML algorithms.
Through a comprehensive experimental evaluation using real-world and synthetic datasets, we compare \dana against the popular in-RDBMS ML toolkit, Apache MADlib~\cite{madlib2}, on both \psql and its parallel counterpart, Greenplum. 
Using Xilinx UltraScale+ VU9P FPGA, we observe \dana generated accelerators provide on average \avgspeeduphcreal and \greenavgspeeduphcreal end-to-end runtime speedups over \psql and Greenplum running MADlib, respectively.
An average of \danastridernstrider of the speedup benefits are obtained through \striders, as they effectively bypass the CPU and its memory subsystem overhead.

\section{Background}

\label{sec:overview}
\begin{figure*}
	\centering
	\includegraphics[width=0.95\linewidth]{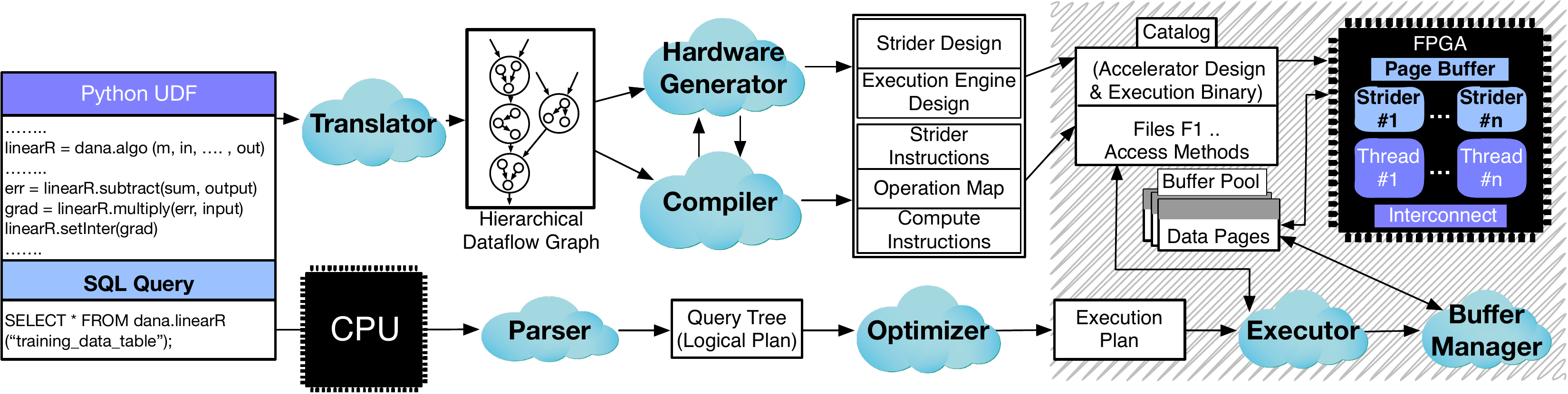}
	\vspace{-1.5ex}
	\caption{Overview of \dana, that integrates FPGA acceleration with the RDBMS engine. The Python-embedded DSL is an interface to express the ML algorithm that is converted to hardware architecture and its execution schedules (stored in the RDBMS catalog). The RDBMS engine fills the buffer pool. FPGA \striders directly access the data pages to extract the tuples and feed them to the threads. Shaded areas show the entangled components of RDBMS and FPGA working in tandem to accelerate in-database analytics.}
	\label{fig:workflow}
	\vspace{-4ex}
\end{figure*}

%Due to their ubiquity across many application domains, machine learning 
%algorithms are constantly evolving to provide intelligent services and more 
%insightful analytics.
%
%The training phase of these algorithms is heavily compute intensive, which can limit their applicability and utility.
%
%To address the increasing demand and computational burden of such algorithms, industry and research communities are moving towards specialized hardware accelerators that provide orders of magnitude higher performance than general-purpose compute platforms.

Before delving into the details of \dana, this section discusses the properties of ML algorithms targeted by our holistic framework. % to provide its holistic framework. 

\subsection{Iterative Optimization and Update Rules}
During training, a wide range of supervised machine learning algorithms go through a cyclic process that demand constant iteration, tuning, and improvement.
%%
%Therefore, one of the most promising solutions for in-database analytics is to describe these algorithms using user-defined functions (UDFs), which iteratively process the data and update the learning model.
%
%These UDFs iteratively process the data through repeated invocation these UDFs and update the learning model.
%%
These algorithms use optimization procedures that iteratively minimize a loss function -- distinct for each learning algorithm -- by using one tuple (input-output pair) at a time to generate updates for the learning model.
%
%Many popular ML algorithms have an iterative nature: they cycle through the training data points and use each point to improve the ML model.
%
Each ML algorithm has a specific loss function that mathematically captures the measure of the learning model's error. 
Improving the model corresponds to minimizing this loss function using an~\emph{update rule}, which is applied repeatedly over the training model, one training data tuple (input-output pair) at a time, until convergence. % by computing over a single or a batch of tuples from the training dataset managed by the database. 

\niparagraph{Example.}
Given a set of $N$ pairs of $\{(x_1,y^*_1),...,(x_N,y^*_N)\}$ constituting the training data, the goal is to find a hypothesis function $h(w^{(t)},x)$ that can accurately map $x \rightarrow y$.
The equation below specifies an entire update rule, where $l(w^{(t)},x_i,y_i^*)$ is the loss function that signifies the error between the output $y^*$ and the predicted output estimated by a hypothesis function $h(w^{(t)},x_i)$ for input $x$.
\begin{equation}
	\label{eq:sgd}
	\displaystyle w^{(t+1)}=w^{(t)} - \mu \times \frac{\partial( l(w^{(t)},x_i,y_i^*) )} {\partial w^{(t)}}
\end{equation}
For each $(x,y^*)$ pair, the goal is to a find a model ($w$) that minimizes the loss function $l(w^{(t)},x_i,y_i)$ using an iterative update rule.
While the hypothesis ($y=h(w,x)$) and loss function vary substantially across different ML algorithms, the optimization algorithm that iteratively minimizes the loss function remains fixed.
%
%As such, the two required components for supervised ML are the loss function and an optimization algorithm.
%
As such, the two required components are the hypothesis function to define the machine learning algorithm and an optimization algorithm that iteratively applies the update rule.

\niparagraph{Amortizing the cost of data accesses by parallelizing the optimization.}
In Equation~(\ref{eq:sgd}), a single ($x_i$, $y_i^*$) tuple is used to update the model.
However, it is feasible to use a batch of tuples and compute multiple updates independently when the optimizer supports combining partial updates~\cite{parallelizability-sgd, parallelized-sgd, minibatched-sgd, slowlearners, distributed-conmaxent, distributed-synchronous-sgd, synchronous-sgd}.
%
%However, using a batch of tuples is feasible when the optimizer supports a linear combination of the partial updates obtained from each tuple~\cite{parallelizability-sgd, parallelized-sgd, minibatched-sgd, slowlearners, distributed-conmaxent, distributed-synchronous-sgd, synchronous-sgd}. 
%
%Often, using a batch of tuples for an analytics function, computes the entire update independently and combines them using a mathematical operator. 
%
This offers a unique opportunity for \dana to rapidly consume data pages brought on-chip by \striders while efficiently utilizing the large, ever-growing amounts of compute resources available on the FPGAs through simultaneous multi-threaded acceleration.
Examples of commonly used iterative optimization algorithms that support parallel iterations are variants of gradient descent methods, which can be applied across a diverse range of ML models. 
%
%These optimization techniques are applied across diverse machine learning models, such as support vector machines, regression models, low rank matrix factorization, least square models, backpropagation, and many more.
%
\dana is equipped to accelerate the training phase of any hypothesis and objective function that can be minimized using such iterative optimization.
%
%The update rule for this family of methods is of the form of Equation \ref{eq:sgd}, where only the gradient of the objective function varies.
%
Thus, the user simply provides the update rule via the \dana DSL described in \S\ref{sec:lang}.
%
%As such, \dana expects the user to provide the update rule via its DSL, which is embedded within Python as a UDF as described in \S\ref{sec:lang}.  
%
%Therefore, the hardware accelerator is not limited to fixed algorithms.
%
%HADI: The following is your text the above is mine
%Examples of commonly used iterative optimization algorithms that support parallel iterations are stochastic gradient descent and variants such as conjugate gradient methods. 
%%
%These optimization techniques are applied across diverse domains of machine learning models, such as support vector machines, logistic regression, linear regression, low rank matrix factorization, least square models, backpropagation, and many more.
%
%We equip \dana to accelerate training any hypothesis function which can be learned using iterative optimization.
%
%Therefore, the hardware accelerator is not bound to fixed algorithms.
%
%Furthermore, \dana expects the user to provide the update rule via its DSL, which is embedded within Python as a UDF as described in \S\ref{sec:lang}.  

In addition to providing a background on properties of machine learning algorithms targeted by \dana, Appendix A in our tech report (\url{http://act-lab.org/artifacts/dana/addendum.pdf}) provides a brief overview on Field Programmable Gate Arrays (FPGAs).
It provides details about the reconfigurability of FPGAs and how they offer a potent solution for hardware acceleration. 

\subsection{Insights Driving \danasec}
\label{sec:insights}
%
%\dana integrates the RDBMS engine and FPGA accelerator to target a wide range of iterative optimization problems by leveraging the following insights.
\niparagraph{Database and hardware interface considerations.}
To obtain large benefits from hardware acceleration, the overheads of a traditional Von-Neumann architecture and memory subsystem need to be avoided.
%the CPU data transformation of the database tables needs to be eliminated. 
%
Moreover, data accesses from the buffer pool need to be at large enough granularities to efficiently utilize the FPGA bandwidth.
\dana satisfies these criteria through \striders, its database-aware reconfigurable memory interface, discussed in \S~\ref{sec:strider}.

\niparagraph{Algorithmic considerations.}
The training data retrieved from the buffer pool and stored on-chip must be consumed promptly to avoid throttling the memory resources on the FPGA. 
\dana achieves this by leveraging the algorithmic properties of iterative optimization to execute multiple instances of the update rule.
The Python-embedded DSL provides a concise means of expressing this update rule for a broad set of algorithms while facilitating parallelization.
%
%Thus, we provide a means for parallelized variants of update rules through our Python-based DSL.
%
%and Python-based DSL needs to allow a broad set of algorithms (focusing on advanced analytics) to be expressed which in turn can be accelerated. 
%
%These insights are derived by \dana, from the algorithmic properties of 
%iterative optimization problems and their flexibility in multi-threading.

\dana leverages these insights to provide a cross-stack solution that generates FPGA-synthesizable accelerators that directly interface with the RDBMS engine's buffer pool. 
The next section provides an overview of \dana.
%integrate with the RDBMS by directly interfacing with RDBMS engine's buffer pool.
\section{\danasec Workflow}
Figure~\ref{fig:workflow} illustrates \dana's integration within the traditional software stack of data management systems.
With \dana, the data scientist specifies her desired ML algorithm as a UDF using a simple DSL integrated within Python. 
\dana performs static analysis and compilation of the Python functions to program the FPGA with a high-performance, energy-efficient hardware accelerator design. 
The hardware design is tailored to both the ML algorithm and page specifications of the RDBMS engine. 
To run the hardware accelerated UDF on her training data, the user provides a SQL query.
\dana stores accelerator metadata (\strider and execution engine instruction schedules) in the RDBMS's catalog along with the name of a UDF to be invoked from the query.
As shown in Figure~\ref{fig:workflow}, the RDBMS catalog is shared by the database engine and the FPGA.
The RDBMS parses, optimizes, and executes the query while treating the UDF as a black box. 
During query execution, the RDBMS fills the buffer pool, from which \dana ships the data pages to the FPGA for processing. 
\dana and the RDBMS engine work in tandem to generate the appropriate data stream, data route, and accelerator design for the \{ML algorithm, database page layout, FPGA\} triad. 
Each component of \dana is briefly described below. 

\niparagraph{Programming interface.}
The front end of \dana exposes a Python-embedded DSL (discussed in \S\ref{sec:lang}) to express the ML algorithm as a UDF. 
The UDF includes an update rule that specifies how each tuple or record in the training data updates the ML model.
%
% by using the operations available in the DSL.
%
It also expects a merge function that specifies how to process multiple tuples in parallel and aggregate the resulting ML models.
\dana's DSL constitutes a diverse set of operations and data types that cater to a wide range of advanced analytics algorithms. % that together form the 
%iterative update rule. 
%
Any legitimate combination of these operations can be automatically converted to a final synthesizable FPGA accelerator.
%
%The details of this Python DSL are discussed in \S\ref{sec:lang}.

\niparagraph{Translator.}
The user provided UDF is converted into a $h$ierarchical DataFlow Graph ($h$DFG) by \dana's parser, discussed in detail in \S\ref{sec:parser}.
Each node in the $h$DFG represents a mathematical operation allowed by the DSL, and each edge is a multi-dimensional vector on which the operations are performed.
The information in the $h$DFG enables \dana's backend to optimally customize the reconfigurable architecture and schedule and map each operation for a high-performance execution.
%
%Each operation in the DSL is converted into a node in $h$DFG and connected so as to maintain the data dependencies according to the UDF.
%
%This $h$DFG also contains auxiliary information such as the dimension of the inputs, outputs and the operation itself.%, for example, scalar-scalar, vector-scalar, matrix-vector, etc. 
%

\niparagraph{Strider-based customizable machine learning architecture.}
To target a wide range of ML algorithms, \dana offers a parametric reconfigurable hardware design solution that is hand optimized by expert hardware designers as described in \S~\ref{sec:template}. 
The hardware interfaces with the database engine through a specialized structure called \striders, that extract high-performance, and provide low-energy computation. 
\striders eliminate CPU from the data transformation process by directly interfacing with database's buffer pool to extract the training data pages. 
They process data at a page granularity to amortize the cost of per-tuple data transfer from memory to the FPGA. 
To exploit this vast amount of data available on-chip, the architecture is equipped with execution engines that run multiple parallel instances of the update rule.  
%
%Additionally, multi-threading provides an avenue to exploit the ever increasing compute resources that are becoming available on the modern FPGAs. 
%
This architecture is customized by \dana's compiler and \hardwaregen in accordance to the FPGA specifications, database page layout, and the analytics function. 

\niparagraph{Instruction Set Architectures.}
Both \striders and the execution engine can be programmed using their  respective Instruction Set Architectures (ISAs). 
The \strider instructions process page headers, tuple headers, and extract the raw training data from a database page.
Different page sizes and page layouts can be targeted using this ISA. 
%
%Additionally, the execution engines come with their independent ISA to run the engine in a selective SIMD mode.
%
%This variation of the SIMD execution amortizes the cost of instruction handling without imposing the traditional shortcomings of SIMD in synchronizing all the processing engines.
%
The execution engine's ISA describes the operation flow required to run the analytics algorithm in selective SIMD mode. 
%
%selectively run the engine in SIMD mode to amortize the cost of instruction handling.
%
%Instructions are statically scheduled to avoid the traditional shortcomings of SIMD execution that require continuous synchronization.% all the processing engines.
%

\niparagraph{Compiler and \hardwaregen.}
\dana's compiler and \hardwaregen ensure compatibility between the $h$DFG and the hardware accelerator.
For the given $h$DFG and FPGA specifications (such as number of DSP Slices and BRAMs), the \hardwaregen determines the parameters for the execution engine and \striders to generate the final FPGA synthesizable accelerator.
The compiler converts the database page configuration into a set of \strider instructions that process the page and tuple headers and transform user data into a floating point format.
Additionally, the compiler generates a static schedule for the accelerator, a map of where each operation is performed, and execution engine instructions.
%
%The number of striders instantiated in the final design depends on the target FPGA resources. %, such as number of BRAMs 
%and BRAM size. 
%
%
%The \hardwaregen~ knits the execution engine and strider design to create the final accelerator that can fit on the FPGA.
%

%\niparagraph{\dana's flowchart.} 
%
%First, the SELECT query is processed on the CPU, where the UDF is treated as a black box by the RDBMS.  
%
%When the SQL statement is executed, data is routed to the FPGA 
%accelerator, where the compatibility between the training data in the table and accelerator is verified.  
%
As described above, providing flexibility and reconfigurability of hardware accelerators for advanced analytics is a challenging but pertinent problem.
\dana is a multifaceted solution that untangles these challenges one by one. %described in the following sections.

\section{Front-End Interface For \danasec}
\dana's DSL provides an entry point for data scientists to exploit hardware acceleration for in-RDBMS advanced analytics.
This section elaborates on the constructs and features of the DSL and how they can be used to train a wide range of learning algorithms for advanced analytics.
This section also explains how a UDF defined in this DSL is translated into an intermediate representation, i.e., in this case a $h$ierarchical DataFlow Graph ($h$DFG).

\subsection{Programming For \danasec}
\label{sec:lang}
\dana exposes a high-level DSL for database users to express their learning algorithm as a UDF.
Embedding this DSL within Python allows support for intricate update rules using a framework familiar to database users whilst not requiring a full language compiler.
This DSL meets the following objectives:
\begin{enumpacked}
\item Incorporates language constructs commonly seen in a wide class of supervised learning algorithms.
\item Supports expression of any iterative update rule, not just variants of gradient descent, whilst conforming to the DSL constructs.
\item Segregates algorithmic specification from hardware-dependent implementation. 
\end{enumpacked}

%We aim to make hardware acceleration available to analysts who have limited knowledge about hardware design.
%
%The analytics algorithm is defined in Python as a special User-Defined Function.
%
%This embedding allows the analysts to use the generate ML model in Python environment.
%
The language constructs of this DSL -- \emph{components}, \emph{data declarations}, \emph{mathematical operations}, and \emph{built-in functions} -- are summarized in Table~\ref{tbl:lang}.
Users express the learning algorithm using these constructs and provide the (1) \emph{update rule} - to decide how each tuple in the training data updates the model; (2) \emph{merge function} - to specify the combination of distinct parallel update rule threads; and (3) \emph{terminator} - to describe convergence.

\subsection{Language Constructs}

\begin{table}
  \centering
  \caption{Language constructs of \dana's Python-embedded DSL.}
  \vspace{-1ex}
  {\includegraphics[width=1\linewidth]{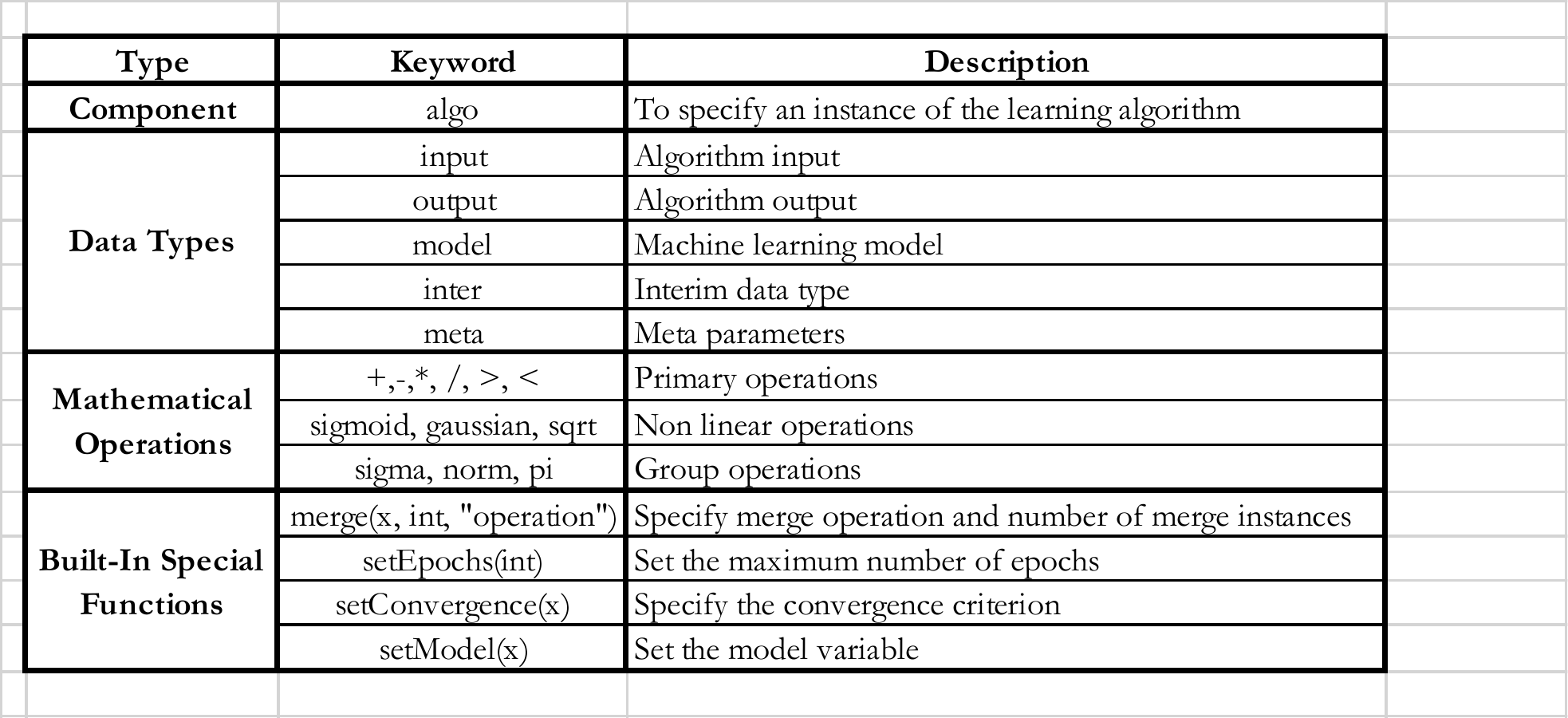}
  \label{tbl:lang}}
  \vspace{-6ex}
\end{table}

\niparagraph{Data declarations.}
Data declarations delineate the semantics of the data types used in the ML algorithm.
The DSL supports the following data declarations: \code{input}, \code{output}, \code{inter}, \code{model}, and \code{meta}.
Each variable can be declared by specifying its type and dimensions.
A variable is an implied scalar if no dimensions are specified.
Once the analyst imports the \code{dana} package, she can express the required variables.
The code snippet below declares a multi-dimensional ML model of size \code{[5][2]} using \code{dana.model} construct. 
\definecolor{black}{rgb}{0,0,0}
\definecolor{grayish}{rgb}{0.85,0.85,0.85}
\definecolor{moregray}{rgb}{0.4,0.4,0.4}

\lstdefinelanguage{danalang}{
	language=python,
	numbers=none,
	stepnumber=2,
	frame=none,
	alsoletter={[*, ), (,<,.,:]},
	escapechar=!,
	aboveskip={0\baselineskip}, 
	basicstyle=\scriptsize\ttfamily,
	keywordstyle={\color{black}\bf\sffamily\bf},
	stringstyle=\sffamily,
	tabsize=2,	
	keywords={},
    emph=[2]{dana.input, dana.output, dana.model, dana.inter, dana.meta, dana.algo, merge, sigma, sigmoid, gaussian, pi, norm, setModel, setVar},
    emphstyle=[2]{\bf\sffamily\bf\uline},
	emph=[3]{readB, extrB, bentr, ad, add, bexit, cln} 
	emphstyle=[3]{\bf\color{moregray}\emph\bf}\lstset
}
\setlength{\fboxsep}{0pt}
%\begin{figure}
\tline
\vspace{1pt}
\begin{lstlisting}[language=danalang, basicstyle=\fontsize{8}{4} \ttfamily]
mo = dana.model ([5][2])
\end{lstlisting}
\vspace{1pt}
\bline
%
%
%, while \code{dana.input} and \code{dana.output} specify the dimensions of a single input-output pair in the training dataset.
%
In addition to \code{dana.model}, the user can provide \code{dana.input} and \code{dana.output} to express a single input-output pair in the training dataset.
The user can specify meta variables using \code{dana.meta}, the value of which remains constant throughout execution.
As such, meta variables can be directly sent to the FPGA before algorithm execution.
All variables used for a particular algorithm are linked to an \code{algo} construct.
\definecolor{black}{rgb}{0,0,0}
\definecolor{grayish}{rgb}{0.85,0.85,0.85}
\definecolor{moregray}{rgb}{0.4,0.4,0.4}

\lstdefinelanguage{danalang}{
	language=python,
	numbers=none,
	stepnumber=2,
	frame=none,
	alsoletter={[*, ), (,<,.,:]},
	escapechar=!,
	aboveskip={0\baselineskip}, 
	basicstyle=\scriptsize\ttfamily,
	keywordstyle={\color{black}\bf\sffamily\bf},
	stringstyle=\sffamily,
	tabsize=2,	
	keywords={},
    emph=[2]{dana.input, dana.output, dana.model, dana.inter, dana.meta, dana.algo, merge, sigma, sigmoid, gaussian, pi, norm, setModel, setVar},
    emphstyle=[2]{\bf\sffamily\bf\uline},
	emph=[3]{readB, extrB, bentr, ad, add, bexit, cln} 
	emphstyle=[3]{\bf\color{moregray}\emph\bf}\lstset
}
\setlength{\fboxsep}{0pt}
%\begin{figure}
\tline
\vspace{1pt}
\begin{lstlisting}[language=danalang, basicstyle=\fontsize{8}{4} \ttfamily]
algorithm = dana.algo (mo, in, out)
\end{lstlisting}
\vspace{1pt}
\bline
The \code{algo} component allows the user to link together the three functions -- update rule, merge, and terminator -- of a single UDF.
%
%Finally the \dana DSL provides \code{dana.iterator} variable to express implicit loops because learning algorithms often require a significant amounts of multi-dimensional operations.
%
Additionally, the analyst can use untyped intermediate variables, which are automatically labeled as  \code{dana.inter} by \dana's backend.
%
%The \code{dana.inter} variables are required by \dana's backend to recognize them as a part of the algorithm.
%

\niparagraph{Mathematical operations.}
The DSL supports mathematical operations performed on both declared and untyped intermediate variables.
Primary and non-linear operations, such as \code{*, +, ... , sigmoid}, only require the operands as input.
The dimensionality of the operation and its output is automatically inferred by \dana's translator (as discussed in \S~\ref{sec:parser}) in accordance to the operands' dimensions.
Group operations, such as \code{sigma, pi, norm}, perform computation across elements.
\code{Sigma} refers to summation, \code{pi} indicates product operator, and \code{norm} calculates the magnitude of a multidimensional vector.
Group operations require the input operands and the grouping axis which is expressed as a constant and alleviates the need to explicitly specify loops.
The underlying primary operation is performed on the input operands prior to grouping.
%
%The dimensionality of any primary operation and resulting intermediate variables is implicitly inferred.

\niparagraph{Built-in functions.}
The DSL provides four built-in functions to specify the merge condition, set the convergence criterion, and link the updated model variable to the \code{algo} component.
The \code{merge(x, int, ``op'')} function is used to specify how multiple threads of the update rule are combined.
Convergence is dictated either by a specifying fixed number of epochs (1 epoch is a single pass over the entire training data set) or a user-specified condition.
Function \code{setEpochs(int)} sets the number of terminating epochs and \code{setConvergence(x)} frames termination based on a boolean variable \code{x}.
Finally, the \code{setModel(x)} function links a \dana variable (the updated model) to the corresponding \code{algo} component.

All the described language constructs are supported by \dana's reconfigurable architecture, hence, can be synthesized on an FPGA.
An example usage of these constructs to express the update rule, merge function, and convergence for linear regression algorithm running the gradient descent optimizer is provided below.

\subsection{Linear Regression Example}
\label{sec:lr-example}

\niparagraph{Update rule.}
%
%This function updates the learning model using a single or a collection of tuples from the database table.
%
As the code snippet below illustrates, the data scientist first declares different data types and their corresponding dimensions. 
Then she defines the computations performed over these variables specific to linear regression.
\definecolor{black}{rgb}{0,0,0}
\definecolor{grayish}{rgb}{0.85,0.85,0.85}
\definecolor{moregray}{rgb}{0.4,0.4,0.4}

\lstdefinelanguage{danalang}{
	language=python,
	numbers=none,
	stepnumber=2,
	frame=none,
	alsoletter={[*, ), (,<,.,:]},
	escapechar=!,
	aboveskip={0\baselineskip}, 
	basicstyle=\scriptsize\ttfamily,
	keywordstyle={\color{black}\bf\sffamily\bf},
	stringstyle=\sffamily,
	tabsize=2,	
	keywords={},
    emph=[2]{dana.input, dana.output, dana.model, dana.inter, dana.meta, dana.algo, merge, sigma, sigmoid, gaussian, pi, norm, setModel, setVar},
    emphstyle=[2]{\bf\sffamily\bf\uline},
	emph=[3]{readB, extrB, bentr, ad, add, bexit, cln} 
	emphstyle=[3]{\bf\color{moregray}\emph\bf}\lstset
}
\setlength{\fboxsep}{0pt}
\tline
\vspace{-3pt}
\begin{lstlisting}[language=danalang, basicstyle=\fontsize{8}{4} \ttfamily]

#Data Declarations
mo = dana.model ([10])
in = dana.input ([10])
out  = dana.output ()
lr = dana.meta (0.3) #learning rate

linearR = dana.algo (mo, in, out)
    
#Gradient or Derivative of the Loss Function
s = sigma ( mo * in, 1)
er = s - out
grad = er * in
 
#Gradient Descent Optimizer
up = lr * grad
mo_up = mo - up
linearR.setModel(mo_up) 
\end{lstlisting}
\vspace{2pt}
\bline

%self.linearR.setGradient(g)
%
In this example, the update rule uses the gradient of the loss function.
The gradient descent optimizer updates the model in the negative direction of the loss function derivative  ($\frac{\partial( l )} {\partial w^{(t)}} $).
The analyst concludes with the \code{setModel()} function to identify the updated model, in this case \code{mo\_up}.
%
%
%After defining these data types, the computations performed over them are defined.
%
%A series of mathematical operations performed over the different data types are then assigned to the algorithm object, as shown below.
%
%The mathematical constructs in this DSL allow a concise definition of a wide range of multi-dimensional operations frequently used in data analytics.
%
%We use a similar set of operations as the domain specific language, developed in our prior work~\cite{tabla}, as it is open source and publicly available (\tablaurl).
%
%All the constructs in the DSL are supported by the reconfigurable architecture, hence can be synthesized on the FPGA. %is equipped to handle all of them.
%

\niparagraph{Merge function.}
The merge function facilitates multiple concurrent threads of the update rule on the FPGA accelerator by specifying the functionality at the point of merge. %mitigate some of the inherent sequential limitations of the algorithm.
%
%For instance, the merge function can combine the output either from the loss function or the optimization function.
%
%The user describes the variable that is to be merged as follows.
%
\definecolor{black}{rgb}{0,0,0}
\definecolor{grayish}{rgb}{0.85,0.85,0.85}
\definecolor{moregray}{rgb}{0.4,0.4,0.4}

\lstdefinelanguage{danalang}{
	language=python,
	numbers=none,
	stepnumber=2,
	frame=none,
	alsoletter={[*, ), (,<,.,:]},
	escapechar=!,
	aboveskip={0\baselineskip}, 
	basicstyle=\scriptsize\ttfamily,
	keywordstyle={\color{black}\bf\sffamily\bf},
	stringstyle=\sffamily,
	tabsize=2,	
	keywords={},
    emph=[2]{dana.input, dana.output, dana.model, dana.inter, dana.meta, dana.algo, merge, sigma, sigmoid, gaussian, pi, norm, setModel, setVar},
    emphstyle=[2]{\bf\sffamily\bf\uline},
	emph=[3]{readB, extrB, bentr, ad, add, bexit, cln} 
	emphstyle=[3]{\bf\color{moregray}\emph\bf}\lstset
}
\setlength{\fboxsep}{0pt}
\tline
\vspace{1pt}
\begin{lstlisting}[language=danalang, basicstyle=\fontsize{8}{4} \ttfamily]
merge_coef = dana.meta (8)
grad = linearR.merge(grad, merge_coef, "+")
\end{lstlisting}
\vspace{1.5pt}
\bline
In the above merge function, the intermediate \code{grad} variable has been combined using addition, and the merge coefficient (\code{merge\_coef}) specifies the batch size.
\dana's compiler implicitly understands that the merge function is performed before the gradient descent optimizer.
Specifically, the \code{grad} variable is calculated separately for each tuple per batch.
The results are aggregated together across the batches and used to update the model.
Alternatively, partial model updates for each batch could be merged.% as shown below.
\definecolor{black}{rgb}{0,0,0}
\definecolor{grayish}{rgb}{0.85,0.85,0.85}
\definecolor{moregray}{rgb}{0.4,0.4,0.4}

\lstdefinelanguage{danalang}{
	language=python,
	numbers=none,
	stepnumber=2,
	frame=none,
	alsoletter={[*, ), (,<,.,:]},
	escapechar=!,
	aboveskip={0\baselineskip}, 
	basicstyle=\scriptsize\ttfamily,
	keywordstyle={\color{black}\bf\sffamily\bf},
	stringstyle=\sffamily,
	tabsize=2,	
	keywords={},
    emph=[2]{dana.input, dana.output, dana.model, dana.inter, dana.meta, dana.algo, merge, sigma, sigmoid, gaussian, pi, norm, setModel, setVar},
    emphstyle=[2]{\bf\sffamily\bf\uline},
	emph=[3]{readB, extrB, bentr, ad, add, bexit, cln} 
	emphstyle=[3]{\bf\color{moregray}\emph\bf}\lstset
}
\setlength{\fboxsep}{0pt}
%\begin{figure}
\tline
\vspace{1pt}
\begin{lstlisting}[language=danalang, basicstyle=\fontsize{8}{4} \ttfamily]
merge_coef = dana.meta (8)
m1 = linearR.merge(mo_up, merge_coef, "+")
m2 = m1/merge_coef
lineaR.setModel(m2)
\end{lstlisting}
\vspace{2pt}
\bline
The \code{mo\_up} is calculated by each thread for tuples in its batch separately and consecutively averaged.
%
%The first merge function specifies the batched gradient descent optimizer, while the second specifies the stochastic gradient descent optimizer.
%
%\todo{Divya}{I think you need to explain the differences between m1 and m2 more clearly.}
%
%These are just two examples of merge descriptions. 
%
Thus, \dana's DSL provides the flexibility to create different learning algorithms without requiring any modification to the update rule by specifying different merge points.
In the above example, the first definition of the merge function creates a linear regression running batched gradient descent optimizer, whereas, the second definition corresponds to a parallelized stochastic gradient descent optimizer.
%
%Merge provides the programmer the flexibility to specify different algorithms without any changes to the update rule function.

\niparagraph{Convergence function.}
The user also provides the termination criteria.
%
%The convergence can also be applied to any variable.
%
As shown in the code snippet below, the convergence checks for the \code{conv} variable, which, if true, terminates the training. 
Variable \code{conv} compares the Euclidean norm of \code{grad} with a \code{conv\_factor} constant.
\definecolor{black}{rgb}{0,0,0}
\definecolor{grayish}{rgb}{0.85,0.85,0.85}
\definecolor{moregray}{rgb}{0.4,0.4,0.4}

\lstdefinelanguage{danalang}{
	language=python,
	numbers=none,
	stepnumber=2,
	frame=none,
	alsoletter={[*, ), (,<,.,:]},
	escapechar=!,
	aboveskip={0\baselineskip}, 
	basicstyle=\scriptsize\ttfamily,
	keywordstyle={\color{black}\bf\sffamily\bf},
	stringstyle=\sffamily,
	tabsize=2,	
	keywords={},
    emph=[2]{dana.input, dana.output, dana.model, dana.inter, dana.meta, dana.algo, merge, sigma, sigmoid, gaussian, pi, norm, setModel, setVar},
    emphstyle=[2]{\bf\sffamily\bf\uline},
	emph=[3]{readB, extrB, bentr, ad, add, bexit, cln} 
	emphstyle=[3]{\bf\color{moregray}\emph\bf}\lstset
}
\setlength{\fboxsep}{0pt}
%\begin{figure}
\tline
\vspace{1pt}
\begin{lstlisting}[language=danalang, basicstyle=\fontsize{8}{4} \ttfamily]
convergenceFactor = dana.meta (0.01)
n = norm(grad , i)
conv = n < convergenceFactor
linear.setConvergence(conv)
\end{lstlisting}
\vspace{2pt}
\bline

Alternatively, the number of epochs can be used for convergence  using the syntax \code{linearR.setEpochs(10000)}.

\niparagraph{Query.}
A UDF comprising the update rule, merge function, and convergence check describes the entire analytics algorithm.
%
%As such, it specifies the analytics algorithm in its entirety for \dana.
%This is all \dana requires to specify the analytics algorithm.
%
The \code{linearR} UDF can then be called within a query as follows:
\definecolor{black}{rgb}{0,0,0}
\definecolor{grayish}{rgb}{0.85,0.85,0.85}
\definecolor{moregray}{rgb}{0.4,0.4,0.4}

\lstdefinelanguage{danalang}{
	language=python,
	numbers=none,
	stepnumber=2,
	frame=none,
	alsoletter={[*, ), (,<,.,:]},
	escapechar=!,
	aboveskip={0\baselineskip}, 
	basicstyle=\scriptsize\ttfamily,
	keywordstyle={\color{black}\bf\sffamily\bf},
	stringstyle=\sffamily,
	tabsize=2,	
	keywords={},
    emph=[2]{dana.input, dana.output, dana.model, dana.inter, dana.meta, dana.algo, merge, sigma, sigmoid, gaussian, pi, norm, setModel, setVar},
    emphstyle=[2]{\bf\sffamily\bf\uline},
	emph=[3]{readB, extrB, bentr, ad, add, bexit, cln} 
	emphstyle=[3]{\bf\color{moregray}\emph\bf}\lstset
}
\setlength{\fboxsep}{0pt}
\tline
\vspace{1pt}
\begin{lstlisting}[language=danalang, basicstyle=\fontsize{8}{5} \ttfamily]
SELECT * FROM dana.linearR('training_data_table');
\end{lstlisting}
\vspace{2pt}
\bline

Currently, for high efficiency and low latency, \dana's DSL and compiler do not support dynamic variables, as the FPGA and CPU do not interchange runtime values and only interact for data handoff. 
\dana only supports variable types which either have been explicitly instantiated as \dana's data declarations, or inferred as intermediate variables (\code{dana.inter}) by \dana's translator.
As such, this Python-embedded DSL provides a high level programming abstraction that can easily be invoked by an SQL query and extended to incorporate algorithmic advancements.
In the next section we discuss the process of converting this UDF into a $h$DFG.

\subsection{Translator}
\label{sec:parser}
\begin{figure}
    \centering
     \vspace{-2ex}
    \subfloat[\sffamily Code snippet]{\includegraphics[width=0.2\textwidth]{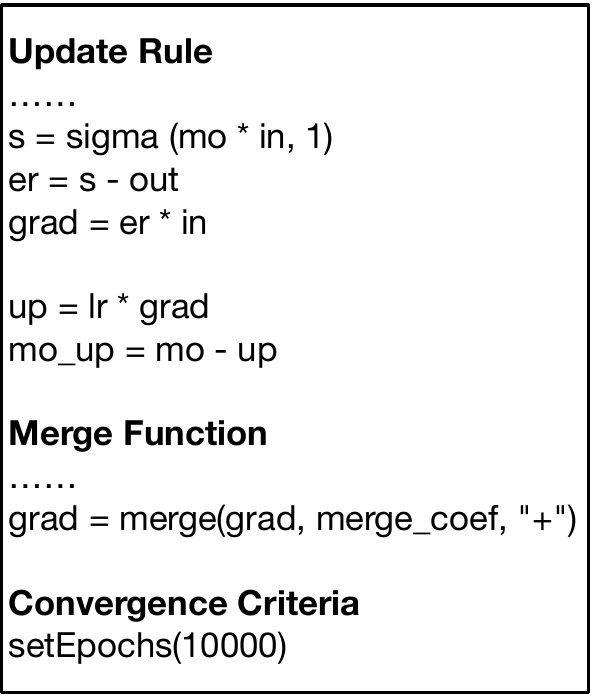}\label{fig:dfg_code}} \quad
    \subfloat[\sffamily Hierarchical DFG]{\includegraphics[width=0.2\textwidth]{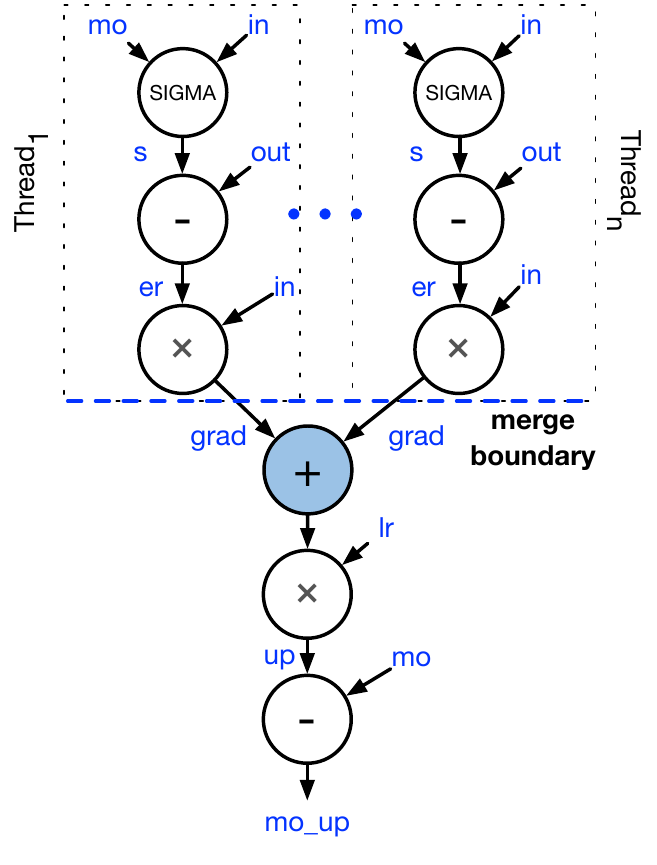}\label{fig:dfg}}    
    \vspace{-2ex}
    \caption{Translator-generated $h$DFG for the linear regression code snippet expressed in \dana's DSL.}
    \label{fig:hdfg}
    \vspace{-2ex}
\end{figure}

%
%The UDF provided by the user is converted to a Hierarchical DataFlow Graph ($h$DFG) by \dana's front end, dubbed translator. 
\dana's translator is the front-end of the compiler, which converts the user-provided UDF to a $h$ierarchical DataFlow Graph ($h$DFG). 
The $h$DFG represents the coalesced update rule, merge function, and convergence check whilst maintaining the data dependencies.
%
%Each node in the $h$DFG constitutes a multi-dimensional operation created by coalescing the update rule, merge function, and convergence check whilst maintaining the data dependencies. 
%
Each node of the $h$DFG represents a multi-dimensional operation, which can be decomposed into smaller atomic sub-nodes.
An atomic sub-node is a single operation performed by the accelerator. 
The $h$DFG transformation for the linear regression example provided in the previous section is shown in Figure~\ref{fig:hdfg}.

%The code snippet and translator-generated $h$DFG for linear regression is shown in Figure~\ref{fig:dfg_code} and Figure~\ref{fig:dfg}, respectively.
%Figure~\ref{fig:dfg_code} and Figure~\ref{fig:dfg} show the code snippet of linear regression and the translator generated $h$DFG , respectively. % update rule and the merge function. 
%
%The figure on the right shows the output generated from the trana. 
%
The aim of the translator is to expose as much parallelism available in the algorithm to the remainder of the \dana workflow.
This includes parallelism within a single instance of the update rule and among different threads, each running a version of the update rule.
To accomplish this, the translator (1) maintains the function boundaries, especially between the merge function and parallelizable portions of the update rule, and (2) automatically infers dimensionality of nodes and edges in the graph. % (3) maintains the data semantics offered by dana i.e. operand type such as input, output, model, meta information is retained in the data flow graph. 

The merge function and convergence criteria are performed once per epoch. 
In Figure~\ref{fig:dfg}, the colored node represents the merge operation that combines the gradients generated by separate instances of the update rule.
These update rule instances are run in parallel and consume different records or tuples from the training data; thus, they can be readily parallelized across multiple threads.
%
%A single instance of the update rule generally suffers from limited parallelism in each node of the DFG due to data dependencies.
%
%For example, the \textbf{sigma} node can be expanded into sub operations, as shown in Table~\ref{fig:pc_isa}.
%
%Even though the initial 8 multiplications can be computed in parallel, the performance of the machine is restricted by the reduction of their products, which exhibits high data dependencies. 
%
%Running multiple instances of the update rule allows the accelerator to exploit the pages of training data available on chip and increases the utilization of the compute resources on the FPGA.
%
To generate the $h$DFG, the translator first infers the dimensions of each operation node and its output edge(s). 
%
%For each node in the  $h$DFG, the translator uses the dimensions of the inputs and the operation type to calculate the output dimension. 
%
For \code{basic operations}, if both the inputs have same dimensions, it translates into an element by element operation in the hardware. 
In case the inputs do not have same dimensions, the input with lower dimension is logically replicated, and the generated output possess the dimensions of the larger input. 
%
%For example, to perform a scalar-vector multiplication, the macro node have one vector input and one scalar input. 
%
%The translator infers the operation and output dimensions to be same as the vector input, where the scalar input is multiplied to each element of the vector. 
%
\code{Nonlinear operations} have a single input that determines the output dimensions.
For \code{group operations}, the output dimension is determined by the axis constant.
%
%is reduced in the same direction as the specified variable across which the grouping has to take place. 
%
For example, a node performing \code{sigma(mo * in, 2)}, where variables \code{mo} and \code{in} are matrices of sizes [5][10] and [2][10], respectively, generates a [5][2] output.

The information captured within the $h$DFG allows the hardware generator to configure the accelerator architecture to optimally cater for its operations.
Resources available on the FPGA are distributed on-demand within and across multiple threads. 
Furthermore, \dana's compiler maps all the operations to the accelerator architecture to exploit fine-grained parallelism within an update rule.
Before delving into the details of hardware generation and compilation, we discuss the reconfigurable architecture for the FPGA (\strider and execution engine).
%
%We elaborate on how the architecture and ISA facilitate the high-performance execution of iterative optimization algorithms.
%
%Using these techniques, the translator generate a comprehensive $h$DFG. 
%
%The $h$DFG created by the translator is then used by the hardware generator and scheduler to generate the final accelerator and its run-time binary.
%

\section{Hardware Design for in-Database Acceleration}
\label{sec:template}

%\todo{Hadi $->$s Divya: everywhere }{access engine $->$ Multi-Threaded Access Engine; execute engine $->$ Multi-Threaded Execute Engine}
%
%\dana provides a reconfigurable architecture that can directly interface with the database to get the training data and perform all the operations in the $h$DFG of the UDF.
%
%This direct integration provides a means to extract the data from the buffer pool of the data, thereby bypassing the CPU and its generic memory subsystem.%, such as
%
%Reconfigurable architecture has the flexibility to execute a range of advanced analytics functions.
%
%Reconfigurability is therefore essential that the hardware accelerator is either programmable or can be reconfigured or both.
%
%As we are targeting FPGAs as the underlying platform, we can reconfigure the design and customize it according to the UDF.
%
%To do so, \dana uses a parametric reconfigurable architecture which, constitutes an \emph{Access Engine} and an \emph{Execution Engine}.
%
\dana employs a parametric accelerator architecture comprising a \emph{multi-threaded access engine} and a \emph{multi-threaded execution engine}, shown in Figure~\ref{fig:accelerator}.
Both engines have their respective custom Instruction Set Architectures (ISA) to program their hardware designs.
%
%Both the access and execution engine come with their custom-made Instruction Set Architecture (ISA), which enables low-level programmability of the accelerator.
%
%This ISA further enables \dana to target the architecture not only for a range of algorithms, but rather different RDBMS engines which share page layout properties, discussed in detail in section~\ref{sec:isa_strider}.
%
The access engine harbors \striders to ensure compatibility between the data stored in a particular database engine and the execution engines that perform the computations required by the learning algorithm.
The access and execution engines are configured according to the page layout and UDF specification, respectively.
The details of each of these components are discussed below.
%The remainder of the section discusses each of these components in greater detail.

\subsection{Access Engine and Striders}
\label{sec:strider}
\subsubsection{Architecture and Design}
\begin{figure*}
\centering
\includegraphics[width=0.75\linewidth]{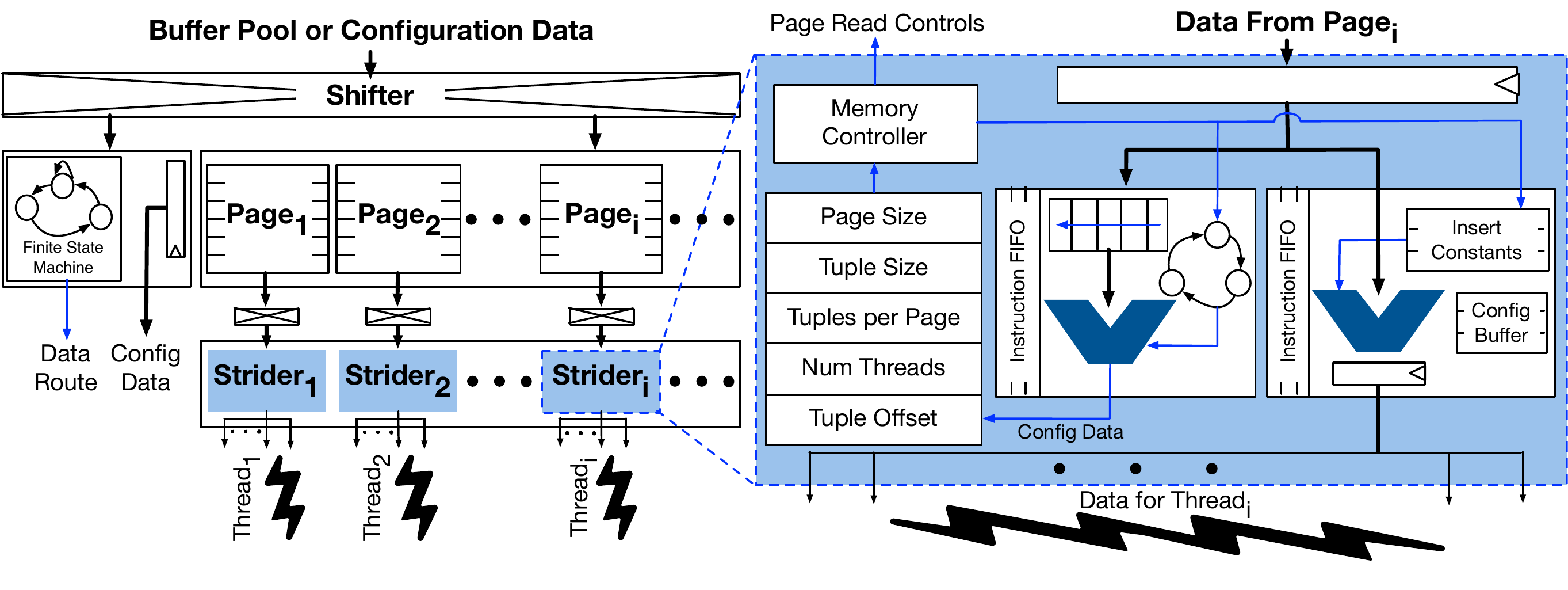}
\vspace{-2ex}
\caption{Access engine design uses \striders as the main interface between the RDBMS and execution engines. Uncompressed data pages are read from the buffer pool and stored in on-chip page buffers. Each page has a corresponding strider to extract the tuple data.}
\vspace{-4ex}
\label{fig:strider}
\end{figure*}

The multi-threaded access engine is responsible for storing pages of data and converting them from a database page format to raw numbers that are processed by the execution engine.
Figure~\ref{fig:strider} shows a detailed diagram of this access engine.
The access engine uses the Advanced Extensible Interface (AXI) interface to transfer the data to and from the FPGA, the shifters properly align the data, and the \striders unpack the database pages.
AXI interface is a type of Advanced Microcontroller Bus Architecture open-standard, on-chip interconnect specification for system-on-a-chip (SoC) designs. 
It is vendor agnostic and standardized across different hardware platforms.
The access engine uses this interface to transfer uncompressed database pages to page buffers and configuration data to configuration registers.
Configuration data comprises \strider and execution engine instructions and necessary meta-data.
Both the training data in the database pages and the configuration data are passed through a shifter for alignment, according to the read width of the block RAM on the target FPGA.
%
%Configuration data comprises strider and execution engine instructions, as well as meta data, and is processed through a separate channel from the training data.
%
A separate channel for configuration data incorporates a finite state machine to dictate the route and destination of the configuration information.
%
%Training data, on the other hand, goes through a shifter for alignment, according to the read width of the block RAM on the target FPGA, and is written to a page buffer.
%
%The access engine can have more than one page buffer, where each buffer stores one database page at a time.  % where each structure has storage for for the data from a single database page and also its read write controls.
%
%Each page buffer has a corresponding strider.% for processing its data.

To amortize the cost of data transfer and avoid the suboptimal usage of the FPGA bandwidth, the access engine and \striders process database data at a page level granularity.
Training data is written to multiple page buffers, where each buffer stores one database page at a time and has access to its personal \strider.
Alternatively, each tuple could have been extracted from the page by the CPU and sent to the FPGA for consumption by the execution engine.
%
%However, one major shortcoming of the latter technique is that it would be unable to exploit the bandwidth available on the FPGA as only one tuple will be sent at a time.
%
This approach would fail to exploit the bandwidth available on the FPGA, as only one tuple would be sent at a time.
Furthermore, using the CPU for data extraction would have a significant overhead due to the handshaking between CPU and FPGA.
Offloading tuple extraction to the accelerator using \striders provides a unique opportunity to dynamically interleave unpacking of data in the access engine and processing it in the execution engine.

\begin{figure}
\centering
\includegraphics[width=0.95\linewidth]{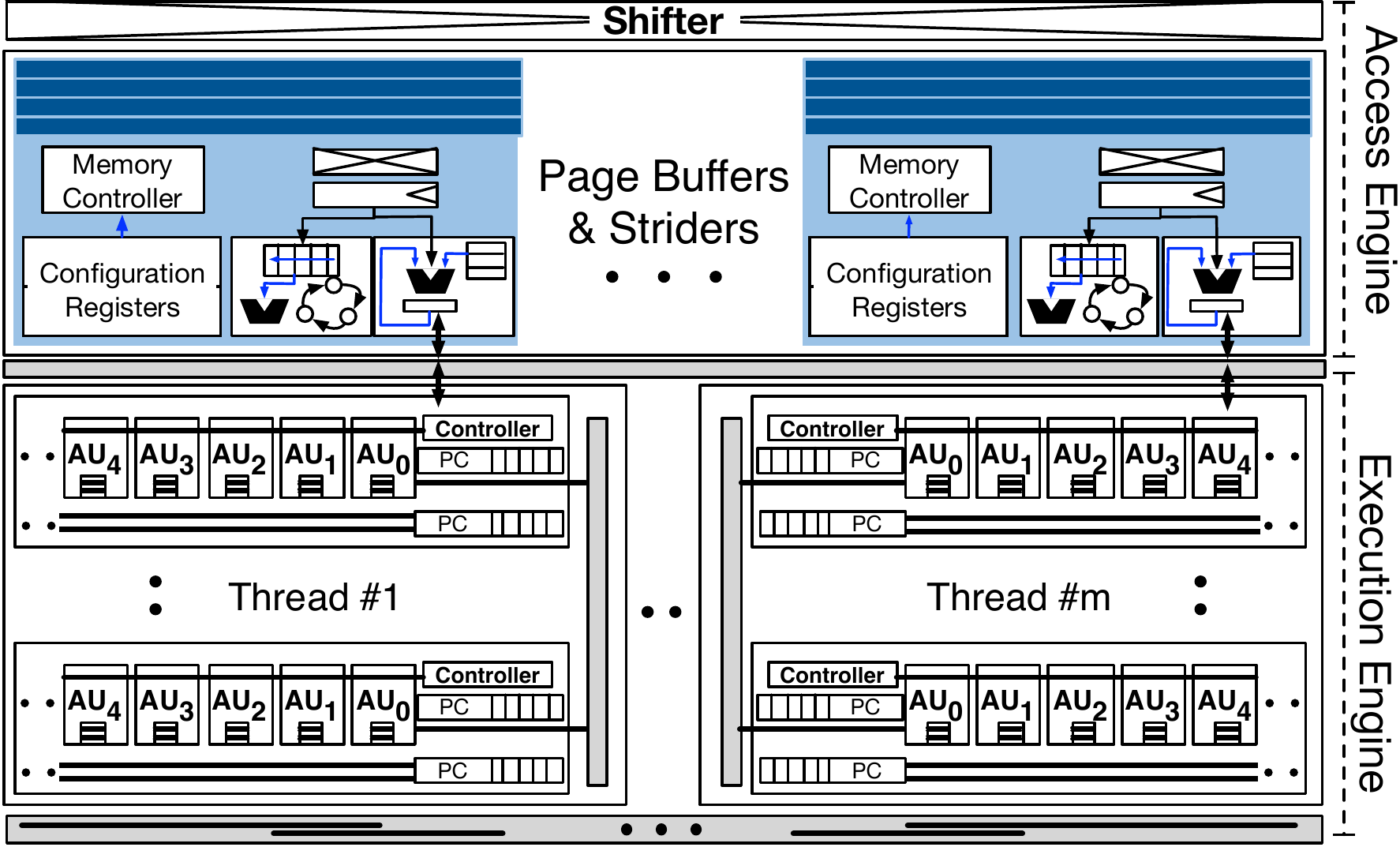}
\vspace{-1.5ex}
\caption{Reconfigurable accelerator design in its entirety. The access engine reads and processes the data via its \striders, while the execution engine operates on this data according to the UDF.}
\vspace{-1.5ex}
\label{fig:accelerator}
\end{figure}

It is common for data to be spread across pages, where each page requires plenty of pointer chasing.
Two tuples cannot be simultaneously processed from a single page buffer, as the location of one could depend on the previous.
Therefore, we store multiple pages on the FPGA and parallelize data extraction from the pages across their corresponding \striders.
%
%Data extraction from the pages is parallelized across their corresponding \striders.
%
%with their corresponding striders to parallelize the process of data extraction.
%
For every page, the \strider first processes the page header and extracts necessary information about the page and stores it in the configuration registers.
The information includes offsets, such as the beginning and size of each tuple, which is either located or computed from the data in the header.
%
%The page header processor is a simple structure with an ALU and small memory.
%
This auxiliary page information is used to trace the tuple addresses and read the corresponding data from the page buffer.
After each page buffer, the shifter ensures alignment of the tuple data for the \strider.
%
%This is vital to efficient performance, as tuple size varies and is dependent on the table schema.
%
%When the tuple is read from the buffer, first the tuple header is processed to calculate the offset for the actual training data.
%
From the tuple data, its header is processed to extract and route the training data to the execution engine.
%
%and the training data is extracted and routed to the execution engine. %, which implements the UDF.
%
The number of \striders and database pages stored on-chip can be adjusted according to the BRAM storage available on the target FPGA.
The internal workings of the \strider are dictated by its instructions that depend on the page layout and page size of the target RDBMS.
We next discuss the novel ISA to program these \striders.

\subsubsection{Instruction Set Architecture for \secstriders}
\label{sec:isa_strider}

\begin{table}
\centering
\caption{\strider ISA to read, extract, and clean the page data.}
\vspace{-1.5ex}
\includegraphics[width=1\linewidth]{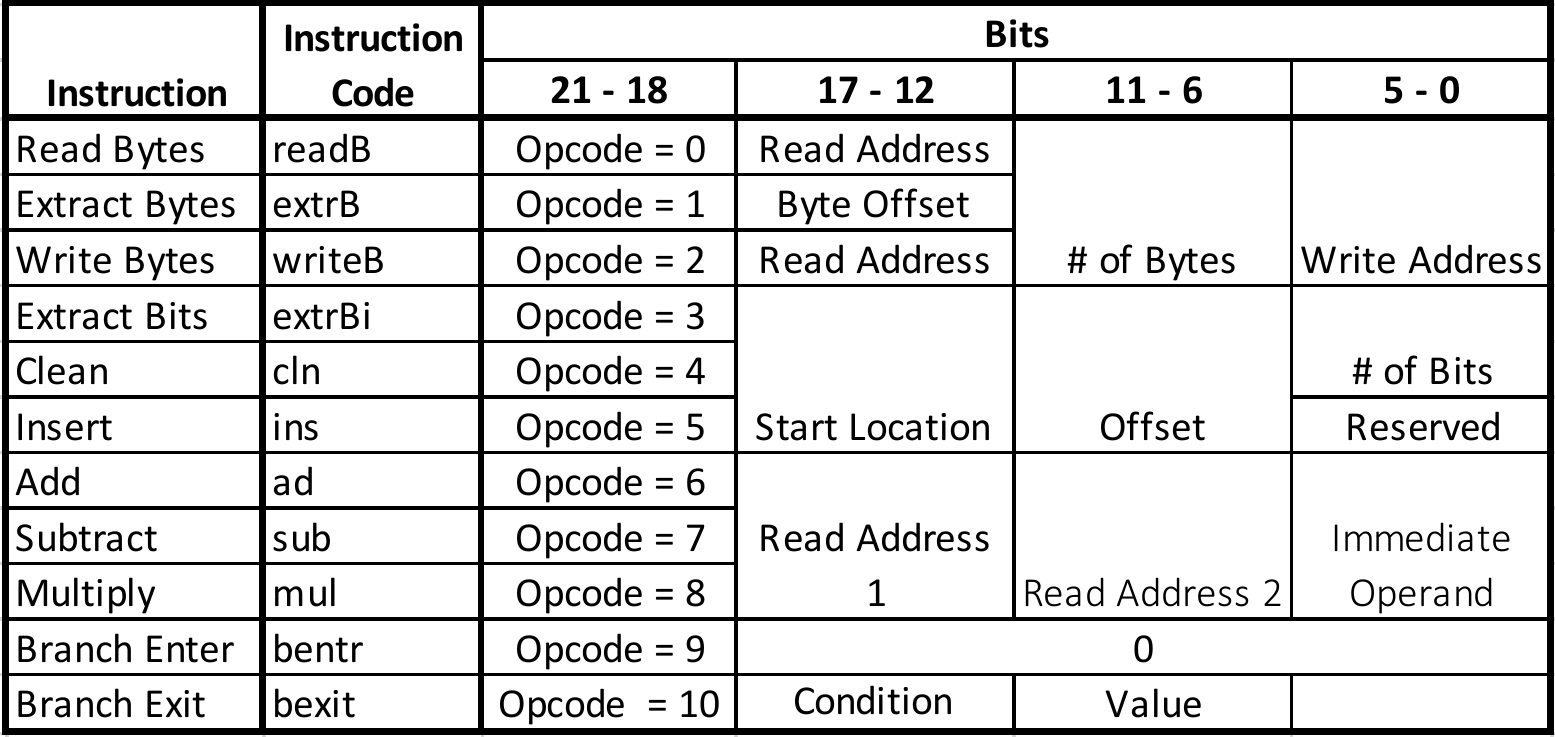}
\label{fig:isa_strider}
\vspace{-6ex}
\end{table}
We devise a novel fixed-length Instruction Set Architecture (ISA) for the \striders that can target a range of RDBMS engines, such as \psql and \mysql (innoDB), that have similar backend page layouts.
An uncompressed page from these RDBMSs, once transferred to the page buffers, can be read, extracted, and cleansed using this ISA, which comprises light-weight instructions specialized for pointer chasing and data extraction.
%, which constitutes majority of the operations.
%
Each \strider is programmed with the same instructions but operates on different pages.
These instructions are generated statically by the compiler.

\begin{figure}
\centering
\includegraphics[width=1\linewidth]{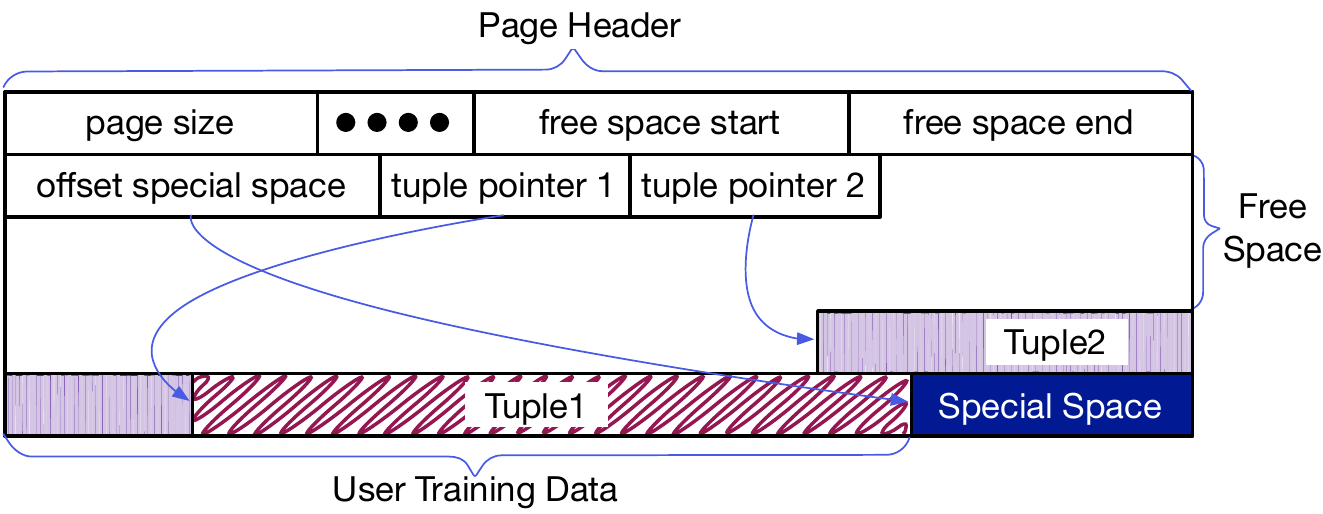}
\vspace{-4.7ex}
\caption{Sample page layout similar to \psql.}
\label{fig:page_layout}
\vspace{-4ex}
\end{figure}

Table~\ref{fig:isa_strider} illustrates the 10 instructions of this ISA.
Every instruction is 22 bits long, comprising a unique operation code (opcode) as identification.
The remaining bits are specific to the opcode.
%
% where each instruction generated is always 20 bits.
%
%The first field is an operation code or the opcode.
%
%Opcode uniquely identifies the instruction and the fields after have a corresponding meaning.
%
Instructions \textbf{Read Bytes} and \textbf{Write Bytes} are responsible for reading and writing data from the page buffer, respectively.
%
%This data is aligned by the shifter and written into a register in the \strider. %, it can be cleaned, enhanced, and computed on by different instructions.
%
%It is common for database pages to be variably aligned.
%
The ISA provides the flexibility to extract data at byte and bit granularity using the \textbf{Extract Byte} and \textbf{Extract Bit} instructions.
%
%Both the extraction instructions operate on the data residing in the \strider.
%
The \textbf{Clean} instruction can remove parts of the data not required by the execution engine.
Conversely, the \textbf{Insert} instruction can add bits to the data, such as NULL characters and auxiliary information, which is particularly useful when the page is to be written back to memory. %in the memory and auxiliary information.
Basic math operations, \textbf{Add}, \textbf{Subtract}, and \textbf{Multiply}, allow calculation of tuple sizes, byte offsets, etc.
Finally, the \textbf{Bentr} and \textbf{Bexit} branch instructions are used to specify jumps or loop exits, respectively.
%
%Bentr and bexit instructions either specify a jump to a previous instruction or lets the strider know when to exit the loop, respectively.
%
This feature invariably reduces the instruction footprint as repeated patterns can be succinctly expressed using branches while enabling loop exits that depend on a dynamic runtime variable.

%The \strider ISA provides the means to extract data from an uncompressed page.
%
An example page layout representative of \psql and \mysql is illustrated in Figure~\ref{fig:page_layout}.
Such layouts are divided into a page header, tuple pointers, and tuple data and can be processed using the following assembly code snippet written in \strider ISA.
\definecolor{black}{rgb}{0,0,0}
\definecolor{grayish}{rgb}{0.85,0.85,0.85}
\definecolor{moregray}{rgb}{0.4,0.4,0.4}

\lstdefinelanguage{danalang}{
	language=python,
	numbers=none,
	stepnumber=2,
	frame=none,
	alsoletter={[*, ), (,<,.,:]},
	escapechar=!,
	aboveskip={0\baselineskip}, 
	basicstyle=\scriptsize\ttfamily,
	keywordstyle={\color{black}\bf\sffamily\bf},
	stringstyle=\sffamily,
	tabsize=2,	
	keywords={},
    emph=[2]{dana.input, dana.output, dana.model, dana.inter, dana.meta, dana.algo, merge, sigma, sigmoid, gaussian, pi, norm, setModel, setVar},
    emphstyle=[2]{\bf\sffamily\bf\uline},
	emph=[3]{readB, extrB, bentr, ad, add, bexit, cln} 
	emphstyle=[3]{\bf\color{moregray}\emph\bf}\lstset
}
\setlength{\fboxsep}{0pt}
\tline
\vspace{1pt}
\begin{lstlisting}[language=danalang, basicstyle=\fontsize{7}{4} \ttfamily]
\\Page Header Processing
readB 0, 8, %cr
readB 8, 2, %cr
readB 10, 4, %cr
extrB %cr, 2, %cr

\\Tuple Pointer Processing
readB %cr, 4, %treg
extrB 0, 1 ,%cr
extrB 1, 1 ,%treg

\\Tuple extraction and processing
bentr
ad %treg, %treg, 0
readB %treg, %cr, %treg 
extrB %treg, %cr, %treg
cln %treg, %cr, 2
bexit 1, %treg, %cr
	
\end{lstlisting} 
\bline
%
%This code is generated by \dana's compiler using the page layout information.
%
Each line in the assembly code is identified by its instruction name (opcode) and its corresponding fields.
%
%This instruction is converted into binary values for the strider controller to comprehend.
%
The first four assembly instructions process the page header to obtain the configuration information.
%such as the page size, page offset from where the tuple space begins and ends, and offset of the special space.
%
For example, the \textbf{(readB 0, 8, \%cr)} instruction, reads 8 bytes from address 0 in the page buffer and adds this page size information into a configuration register.
%
%Similarly, information regarding the tuple space, free space and special/reserved space in the page is also added to other configuration registers.
%
Each variable shown at \%(reg) corresponds to an actual \strider hardware register.
The \%cr is a configuration register, and \%t is a temporary register.
%
%More than one configuration and temporary registers can be read and written.
%
Next, the first tuple pointer is read to extract the byte-offset and length (bytes) of the tuple.
Only the first tuple pointer is processed, as all the training data tuples are expected to be identical. %and the information is retained.
Each corresponding tuple is processed by adding the tuple size to the previous offset to generate the page address.
This address is used to read the data from the page buffer, which is then cleansed by removing its auxiliary information.
The above step is repeated for each tuple using the \textbf{bentr} and \textbf{bexit} instructions.
The loop is exited when the tuple offset address reaches the free space in the page.
Finally, cleaned data is sent to the execution engines.
%

%As discussed above, these instructions are hard-coded into the \striders via the configuration lane of the access engine.
%
%The compiler uses this ISA to customize runtime operations of the \striders to cater for different page sizes and minor differences in the page layout.
%
%In the next section we discusses the execution engine architecture and its ISA.

\subsection{Execution Engine Architecture}
%
%\subsubsection{Architecture and Design}
%
%Processing by \striders, the raw data is transferred to the execution engines to perform the $h$DFG generated from the user provided UDF.
%
The execution engines execute the $h$DFG of the user provided UDF using the \strider processed training data pages.
%
%As discussed above, each strider processes at an uncompressed database page granularity.
%
More and more database pages can now be stored on-chip as the BRAM capacity is  rapidly increasing with the new FPGAs such as Arria~10 that offers 7~MB and UltraScale+~VU9P with 44~MB of memory. 
Therefore, the execution engine needs to furnish enough computational resources that can process this copious amount of on-chip data.
%
%As mentioned before, iterative optimization techniques generally offer the capability to run multiple independent instances of update rule where the result from each can be combined.
%
Our reconfigurable execution engine architecture can run multiple threads of parallel update rules for different data tuples.
This architecture is backed by a \emph{Variable Length Selective SIMD ISA}, that aims to exploit both regular and irregular parallelism in ML algorithms whilst providing the flexibility to each component of the architecture to run independently.
%
%This selective SIMD ISA targets two types of nodes in the $h$DFG, ones that are easily vectorized and ones which exhibit limited parallelism by showing high data dependencies.
%

\niparagraph{Reconfigurable compute architecture.}
All the threads in the execution engine are architecturally identical and perform the same computations on different training data tuples.
\dana balances the resources allocated per thread vs. the number of threads to ensure high performance for each algorithm.
The hardware generator of \dana (discussed in \S\ref{sec:hg}) determines this division by taking into account the parallelism in the $h$DFG, number of compute resources available on chip, and number of striders/page buffers that can fit on the on-chip BRAM.
The architecture of a single thread is a hierarchical design comprising analytic clusters (ACs) composed of multiple analytic units (AUs).
%
%Instead of concentrating of making each thread very high performance and inculcating large area overheads, the hardware generator focuses on delicately balancing the architecture so as to provide high throughput and not throttling the performance due to the sequential nature of a single threaded $h$DFG execution.
%
%Operations within a $h$DFG, that need to performed sequentially due to data dependencies  form the bottleneck of the entire update rule computation.
%
%Multiple threads maybe not necessarily alleviate this bottleneck, but enabling parallel executions can increase throughput.
%
As discussed below, the AC architecture is designed while keeping in mind the algorithmic properties of multi-threaded iterative optimizations, and the AU caters to commonly seen compute operations in data analytics.
%
%The design of AC and AU are discussed below.

%A specialized ISA defines the programmability of this entire architecture.
%
%This execution engine architecture supports a variable length selective SIMD ISA that has been designed to reduce the instruction footprint on the architecture, so that the memory resources can be maximally diverted towards storing the training data.

\begin{figure}
  \centering
   \subfloat[\sffamily Analytic Cluster]{\includegraphics[width=0.9\linewidth]{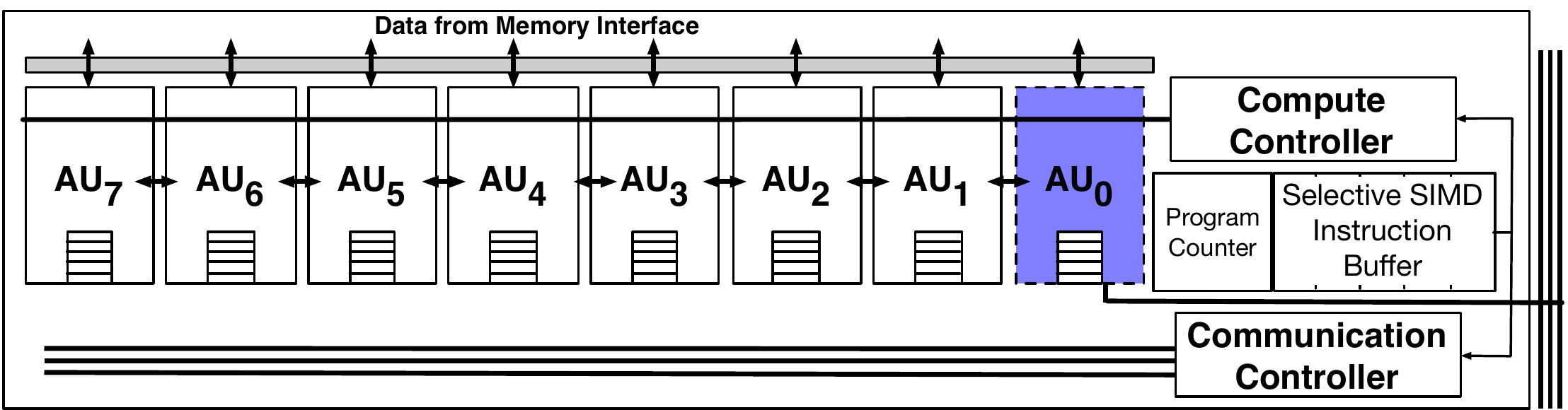}
   \label{fig:ac}}\\
    \vspace{-3ex}
    \subfloat[\sffamily Analytic Unit]{\includegraphics[width=0.8\linewidth]{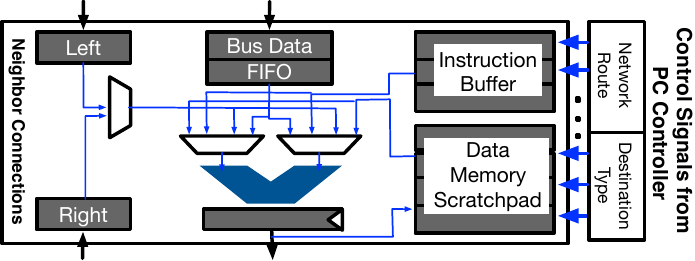}\label{fig:au}}
    \vspace{-2ex}
  \caption{(a) Single analytic cluster comprising analytic units operating in a selective SIMD mode and (b) an analytic unit that is the pipelined compute hub of the architecture.}
  \vspace{-4ex}
\end{figure}

\niparagraph{Analytic cluster.}
An Analytic Cluster (AC), shown in Figure~\ref{fig:ac}, is a collection of AUs designed to reduce the data transfer latency between them.
Thus, $h$DFG nodes which exhibit high data dependencies are all scheduled to a single cluster.
In addition to providing greater connectivity among the AUs within an AC, the cluster serves as the control hub for all its constituent AUs.
%
%As the architecture, the instructions by the code generator (discussed in section~\ref{sec:backend}) are at AC level.
%
The AC runs in a selective SIMD mode, where the AC specifies which AUs within a cluster perform an operation.
Each AU within a cluster is expected to execute either a cluster level instruction (add, subtract, multiply, divide, etc.) or a no-operation (NOP).
%
%The NOP is used to specify that the AU should not perform an operation at the current cycle.
%
Finer details about the source type, source operands, and destination type can be stored in each individual AU for additional flexibility.
%
%As such, the cluster runs in a selective SIMD mode, where AC specifies the AUs which perform the operation, whereas the AU individually has the flexibility to specify the source and destination.
%
%Parts of the instruction can be specific to each PE, however, the type of operation being run is fixed across all the PEs in the PC.
%
%This is done so to reduce the instruction footprint across the accelerator and instead use as much of the memory resource available on the FPGA for training data storage.
%
This collective instruction technique simplifies the AU design, as each AU no longer requires a separate controller to decode and process the instruction.
Instead, the AC controller processes the instruction and sends control signals to all the AUs.
%
%If the PE is executing the instruction, the control signals are heeded else ignored.
%
When the designated AUs complete their execution, the AC proceeds to the next instruction by incrementing the program counter.
To exploit the data locality among the operations performed within an AC, different connectivity options are provided.
Each AU within an AC is connected to both its neighbors, and the AC has a shared line topology bus.
%
%Only one AU can access this bus at a time.
%
The number of AUs per AC are fixed to 8 to obtain highest operational frequency.
A single thread generally contains more than one instance of an AC, each performing instructions independently.
%
%The number of clusters instantiated per thread depend on the number of resources available on the FPGA and the requirements of the $h$DFG.% of the UDA's dataflow graph.
%
%Every AC performs its instruction execution independent of each other.
%
Data sharing among ACs is made possible via a shared line topology inter-AC bus.
%
%Only one AU per AC has access to this inter-AC bus, called the \emph{staple} AU.
%
%If any other AU requires bus access, the data first needs to be sent to the staple AU.
%
%A more complicated network of intra-PC and inter-PC bus would enable higher connectivity among PEs and PCs, however, would also require more FPGA resources.
%
%Next, we discuss the internals of the analytic unit and how it performs these low level mathematical operations.
%

%

\niparagraph{Analytic unit.}
The Analytic Unit (AU) shown in Figure~\ref{fig:au}, is the basic compute element of the execution engine.
It is tailored by the hardware generator to satisfy the mathematical requirements of the $h$DFG.
%
%Preliminary control signals generated from the AC instructions are fed to the AU. % from the PC controller.
%
%The data for each operation can be read from the data memory, which stores the training data and other intermediate results, depending on the source type specified by the instruction.
%
Control signals are provided by the AC.
Data for each operation can be read from the memory according to the source type of the AC instruction.
Training data and intermediate results are stored in the data memory.
%
%The data memory is reserved for the training data and the model whereas the scratchpad stores all the intermediate results.
%
%The memory buffer inside the AU is created from the on-chip BRAMs and generally has more than one read ports to enable a single instruction to read multiple data from the same BRAM.
%
%The memories are written in a FIFO manner, however are read using an address.
%
%A FIFO interface for writing makes the write logic less complicated.
%
Additionally, data can be read from the bus FIFO (First In First Out) and/or the registers corresponding to the left and right neighbor AUs.
%
%As the name suggests, for the bus FIFO, the data is read and written in a FIFO manner.
%
%If the data is needed for more than one instruction, the bus data is written back to the memory for reuse.
%
%The neighbor registers, on the other hand, are always rewritten by a valid output generated by the neighbor AUs.
%
%For each instruction, the required data is read and sent to the ALU.
%
Data is then sent to the Arithmetic Logic Unit (ALU), that executes both basic mathematical operations and complicated non-linear operations, such as sigmoid, gaussian, and square root.
%
%These non-linear operations are implemented as look up tables in the template.\
%
%Basic operations use the on-chip DSP, while the remaining operations are executed on the hardware designed using the FPGA LUTs.
%
The internals of the ALU are reconfigured according to the operations required by the $h$DFG.
The ALU then sends its output to the neighboring AUs, the shared bus within the AC, and/or the memory as per the instruction.

\niparagraph{Bringing the Execution Engine together.}
%
%Both the AC and the AU architectures are highly customizable according to the operations required by the $h$DFG.
%
%Furthermore, the number of resources dedicated to each thread is a design space exploration and decided according the number of parallel operations in the UDA.
%
%Each thread design is identical and runs the exact same instructions on different tuple data.
%
Results across the threads are combined via a computationally-enabled tree bus in accordance to the merge function.
This bus has attached ALUs to perform computations on in-flight data.
%
%This is particularly useful for aggregating results from all the threads.
%
The pliability of the architecture enables \dana to generate high-performance designs that efficiently utilize the resources on the FPGA for the given RDBMS engine and algorithm.
The execution engine is programmed using its own novel ISA.
Due to space constraints, the details of this ISA are added to the Appendix B of our tech report (\url{http://act-lab.org/artifacts/dana/addendum.pdf}). 

\section{Backend for \danasec}
\label{sec:stack}
%
%dana offers a complete compute stack, from a high-level programming interface to a high-performance architecture supported by its separate ISAs.
%
\dana's translator, scheduler, and hardware generator together configure the accelerator design for the UDF and create its runtime schedule.
As discussed in \S~\ref{sec:parser}, the translator converts the user-provided UDF, merge function, and convergence criteria into a $h$DFG.
%
%Each of these nodes in $h$DFG can be fragmented into sub-nodes.
%
%For instance, the \textbf{sigma} node shown in Table~\ref{fig:pc_isa} is expressed as a single node in the $h$DFG, which is comprised of 15 sub-nodes.
%
Each node of the $h$DFG comprises of sub-nodes, where each sub-node is a single instruction in the execution engine. 
Thus, all the sub-nodes in the $h$DFG are scheduled and mapped to the final accelerator hardware design.
%
%The final accelerator hardware design needs to be scheduled and mapped cater for all the sub-nodes in the $h$DFG.
%
The hardware generator outputs a single-threaded architecture for the operations of these sub-nodes and determines the number of threads to be instantiated.
The scheduler then statically maps all operations to this architecture.

\subsection{Hardware Generator}
\label{sec:hg}

The hardware generator finalizes the parameters of the reconfigurable architecture  (Figure~\ref{fig:accelerator}) for the \striders and the execution engine.
The hardware generator obtains the database page layout information, model, and training data schema from the DBMS catalog.
FPGA-specific information, such as the number of DSP slices, the number of BRAMs, the capacity of each BRAM, the number of read/write ports on a BRAM, and the off-chip communication bandwidth are provided by the user.
Using this information, the hardware generator distributes the resources among access and execution engine.
Sizes of the DBMS page, model, and a single training data record determine the amount of memory utilized by each \strider.
Specifically, a portion of the BRAM is allocated to store the extracted raw training data and model.
The remainder of the BRAM memory is assigned to the page buffer to store as many pages as possible to maximize the off-chip bandwidth utilization.

Once the number of resident pages is determined, the hardware generator uses the FPGA's DSP information to calculate the number of AUs which can be synthesized on the target FPGA.
Within each AU, the ALU is customized to contain all the operations required by the $h$DFG.
%
%Only those operations that are used by the DFG are instantiated on the architecture.
%
The number of AUs determines the number of ACs.
%
%The hardware generator always uses all the AUs possible to create the final architecture.
%
Each thread is allocated a number of ACs determined by the merge coefficient provided by the programmer.
It creates at most as many threads as the coefficient.
To decide the allocation of resources to each thread vs. number of threads, we equip the hardware generator with a performance estimation tool that uses the static schedule of the operations for each design point to estimate its relative performance. 
It chooses the smallest and best-performing design point which strikes a balance between the number of cycles for data processing and transfer. 
Performance estimation is viable, as the $h$DFG does not change, there is no hardware managed cache, and the accelerator architecture is fixed during execution. 
Thus, there are no dynamic irregularities that hinder estimation. 
This technique is commensurate with prior works~\cite{tabla:hpca, cosmic:micro, dnnweaver:micro} that perform a similar restricted design space exploration in less than five minutes with estimates within 5\% of the physical measurements.
%
%However, the number of ACs allocated to each thread is determined by the merge coefficient provided by the programmer and creates at-most as many threads as the coefficient.
%
%The ACs dedicated to a thread are catered to by a single \strider, that feeds them the raw training data. % that can be processed by the execution engines.
%

Using these specifications, the hardware generator converts the final architecture into a functional and synthesizable design that can efficiently run the analytics algorithm.

\subsection{Compiler}
\label{sec:compiler}

%Each node in the $h$DFG can be subdivided into its corresponding sub-nodes, where each sub-node is a basic maths operation.
%
The compiler schedules, maps, and generates the micro-instructions for both ACs and AUs for each sub-node in the $h$DFG.
%
%The compilation workflow first schedules all the nodes in the .
%
For scheduling and mapping a node, the compiler keeps track of the sequence of scheduled nodes assigned to each AC and AU on a per-cycle basis.
For each node which is ``ready'', i.e., all its predecessors have been scheduled, the compiler tries to place that operation with the goal to improve throughput.
%
%Lower communication cost is preferable.
%
%Often, a single node in the $h$DFG has dimensions greater than the number of AUs in an AC.
%
%In this case, operations within a node are spread across multiple ACs.
%
Elementary and non-linear operation nodes are spread across as many AUs as required by the dimensionality of the operation.
As these operations are completely parallel and do not have any data dependencies within a node, they can be dispersed.
For instance, an element-wise vector-vector multiplication, where each vector contains 16 scalar values will be scheduled across two ACs (8 AUs per ACs).
Group operations exhibit data dependencies, hence, they are mapped to minimize the communication cost.
%
%The following options are in an increasing order of communication cost, sub-node assigned to the same AU as its predecessor sub-node, sub-node assigned to a neighbor AU , sub-node assigned to a neighbor AU, sub-node assigned non-neighbor AU within the same AC, and operation assigned to a different AC.
 %
%Provided the assignment of a predecessor sub-node, its successors are assigned to the least communication cost AU.
%
%If that AU is already occupied by another operation in the same cycle, the next option is accepted and so on.
%
After all the sub-nodes are mapped, the compiler generates the AC and AU micro-instructions.

The FPGA design, its schedule, operation map, and instructions are then stored in the RDBMS catalog.
These components are executed when the query calls for the corresponding UDF.

\begin{table}
	\centering
	\caption{Descriptions of datasets and machine learning models used for evaluation. Shaded rows are synthetic datasets.}
	\vspace{-2ex}
	\label{tbl:datasets}
	\includegraphics[width=1\linewidth]{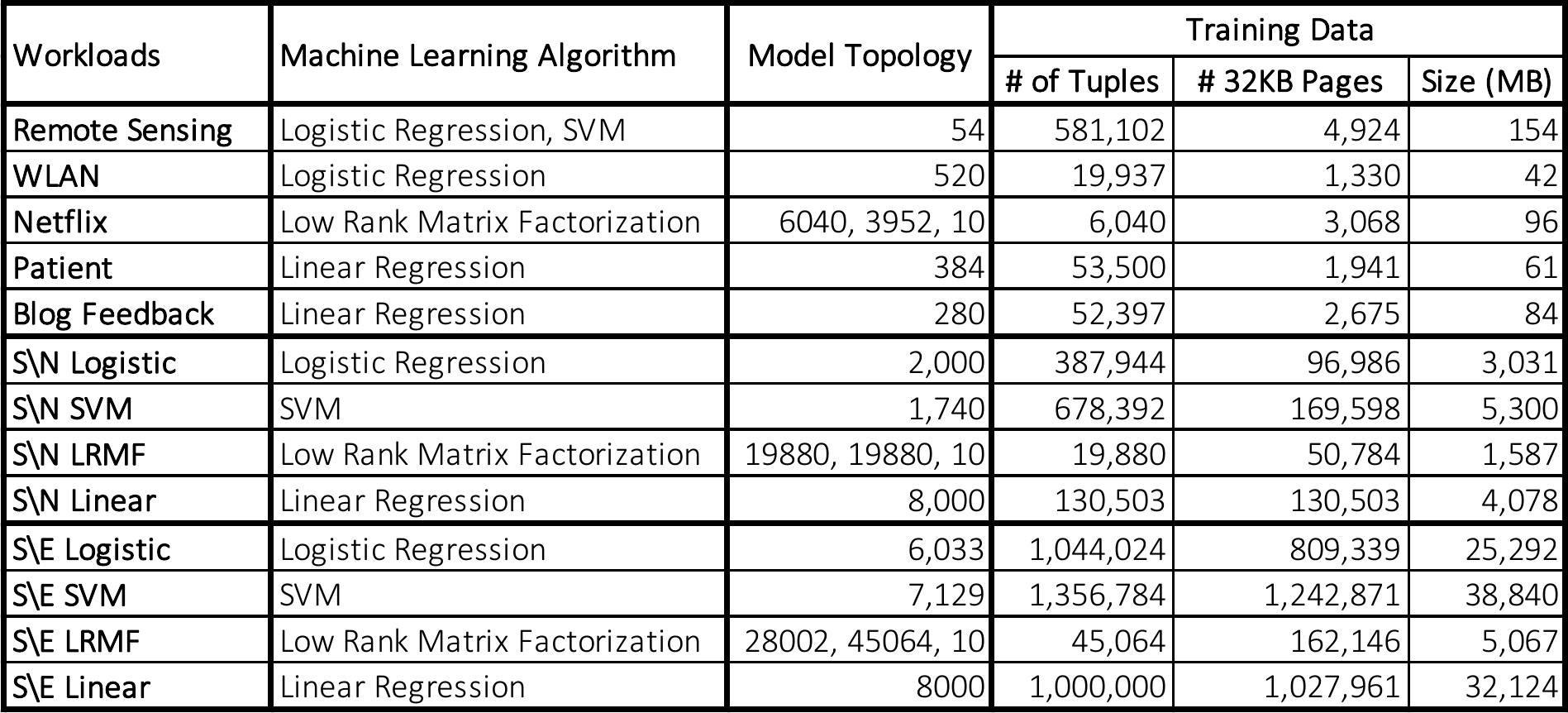}
	\vspace{-4ex}
\end{table}

\begin{table}
	\centering
	\caption{ Xilinx Virtex UltraScale+ VU9P FPGA specifications.}
	\vspace{-2ex}
	\label{tbl:fpga_specs}
	\includegraphics[width=0.9\linewidth]{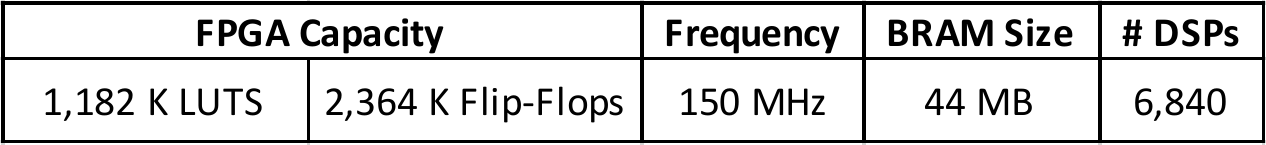}
	\vspace{-4ex}
\end{table}

\section{Evaluation}
\label{sec:eval}

We prototype \dana by integrating it with \psql and compare the end-to-end runtime performance of \dana generated accelerators with a popular scalable in-database advanced analytics library, Apache MADlib~\cite{madlib1, madlib2}, for both \psql and Greenplum RDBMSs. 
We compare the end-to-end runtime performance of these three systems. 
Next, we investigate the impact of \striders on the overall runtime of \dana and how accelerator performance varies with the system parameters. %directly interfacing the hardware accelerators, via striders, with the RDBMS engine. 
Such parameters include the buffer page-size, number of Greenplum segments, multi-threading on the hardware accelerator, and bandwidth and compute capability of the target FPGA.
%
%\revision{The aim is to discern the impact of different  on performance by using publicly available and synthetic datasets for a diverse set of machine learning models.} 
%
We also aim to understand the overheads of performing analytics within RDBMS, thus compare \mpsql with software libraries optimized to perform analytics outside the database.
Furthermore, to delineate the overhead of reconfigurable architecture, we compare our FPGA designs with custom hard coded hardware designs targeting a single or fixed set of machine learning algorithms. 

\niparagraph{Datasets and workloads.}
%
%We use real and synthetic datasets to evaluate \dana.
%
Table \ref{tbl:datasets} lists the datasets and machine learning models used to evaluate \dana. 
These workloads cover a diverse range of machine learning algorithms, -- Logistic Regression (Logistic), Support Vector Machines (SVM), Low Rank Matrix Factorization (LRMF), and Linear Regression (Linear). 
%
%The top part of the table enumerates the specifications of the real datasets. 
%
\bench{Remote Sensing}, \bench{WLAN}, \bench{Patient}, and \bench{Blog Feedback} are publicly available datasets, obtained from the UCI repository~\cite{uci}. 
\bench{Remote Sensing} is a classification dataset used by both logistic regression and support vector machine algorithms. % reherte.
\bench{Netflix} is a movie recommendation dataset for LRMF algorithm. 
The model topology, number of tuples, and number of uncompressed 32~KB pages that fit the entire training dataset are also listed in the table.
Publicly available datasets fit entirely in the buffer pool, hence impose low I/O overheads. 
To evaluate the performance of out-of-memory workloads, we generate eight synthetic datasets, shown by the shaded rows in Table~\ref{tbl:datasets}. 
Synthetic Nominal (S\textbackslash N) and Synthetic Extensive (S\textbackslash E) datasets are used to evaluate performance with the increasing sizes of datasets and model topologies.
Finally, Table~\ref{tbl:runtimes} provides absolute runtimes for all workloads across our three systems.

\niparagraph{Experimental setup.}
We use the Xilinx Virtex UltraScale+ VU9P as the FPGA platform for \dana and synthesize the hardware at 150~MHz using Vivado 2018.2.
Specifications of the FPGA board are provided in Table~\ref{tbl:fpga_specs}. 
\dana accelerators . 
%
%The operational frequency of the FPGA accelerators is \code{150 MHz}. 
%
The baseline experiments for MADlib were performed on a machine with four Intel i7-6700 cores at 3.40GHz running Ubuntu 16.04 xLTS with kernel 4.8.0-41, 32GB memory, a 256GB Solid State Drive storage. 
We run each workload with MADlib v1.12 on \psql v9.6.1 and Greenplum v5.1.0 to measure single- and multi-threaded performance, respectively. 
%
%In addition, we break down the runtime into UDF preprocessing time, compute time, and disk access time.

\niparagraph{Default setup.}
Our default setup uses a 32~KB buffer page size and 8~GB buffer pool size across all the systems. 
As \dana operates with uncompressed pages to avoid on-chip decompression overheads, 32~KB pages are used as a default to fit at least 1 tuple per page for all the datasets. 
To understand the performance sensitivity by varying the page size on \psql and Greenplum, we measured end-to-end runtimes for 8, 16, and 32~KB page sizes. 
We found that page size had no significant impact on the runtimes.
Additionally, we did a sweep for 4, 8, and 16 segments for Greenplum.
We observed the most benefits with 8 segments, making it our default choice. 
Results are obtained for both warm cache and cold cache settings to better interpret the impact of I/O on the overall runtime. 
In the case of a warm cache, before query execution, training data tables for the publicly available dataset reside in the buffer pool, whereas only a part of the synthetic datasets are contained in the buffer pool.  
For the cold cache setting, before execution, no training data tables reside in the buffer pool. 

\begin{table}
	\centering
	\caption{Absolute runtimes across all systems.}
	\vspace{-2ex}
	\label{tbl:runtimes}
	\includegraphics[width=0.9\linewidth]{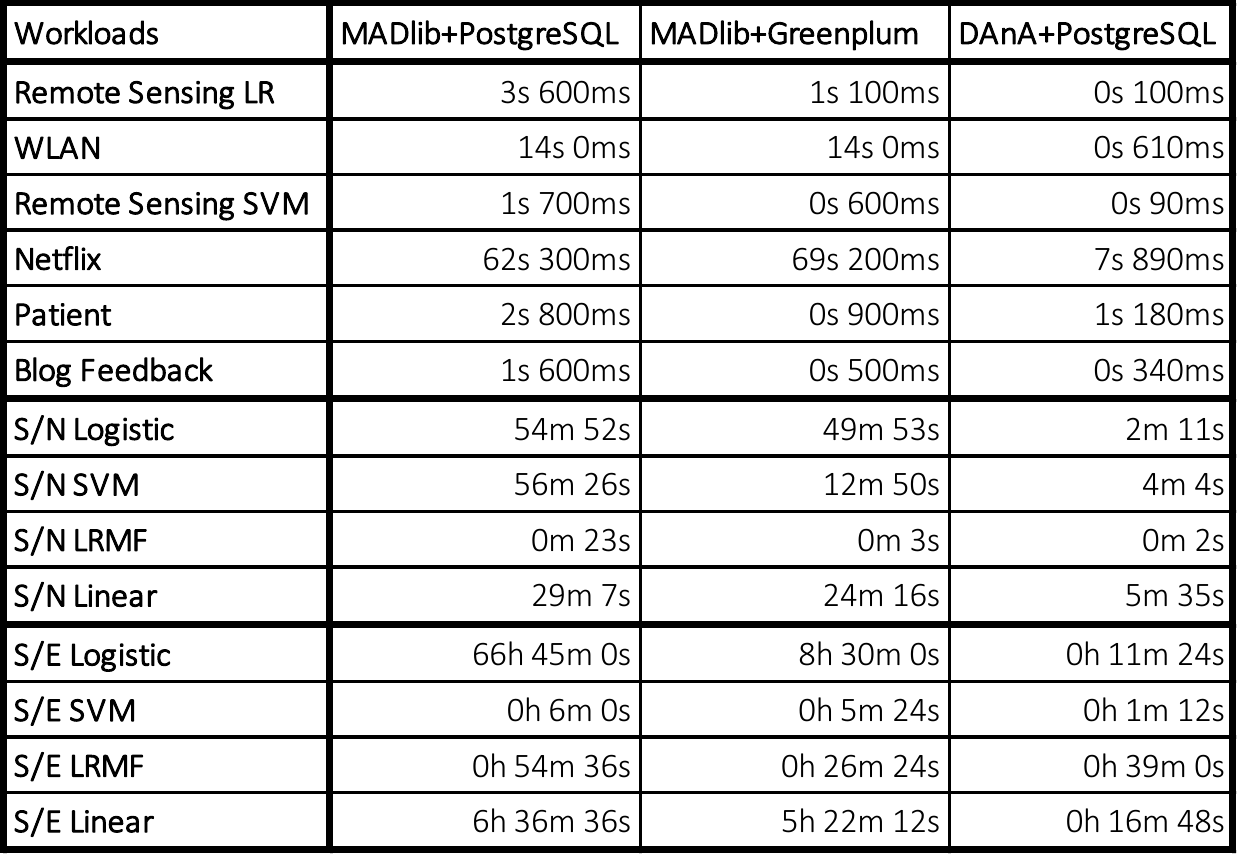} 
	\vspace{-4ex}
\end{table}

\subsection{End-to-End Performance}
\label{sec:eteperf}

\begin{figure*}[ht]
  \centering
  \captionsetup[subfloat]{captionskip=1.6pt}
   \subfloat[\sffamily End-to-End Performance for Warm Cache]{\includegraphics[width=0.48\linewidth]{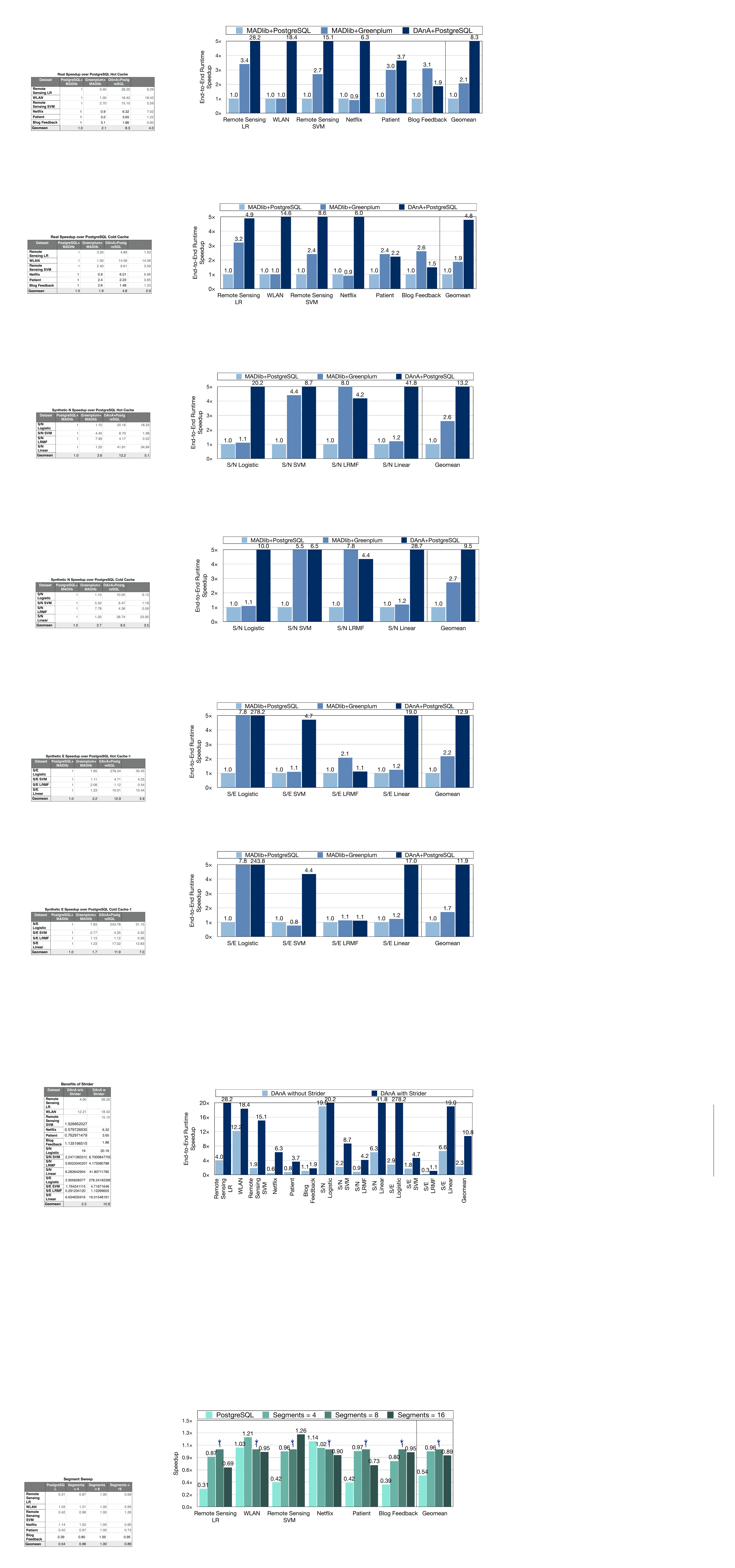} \label{fig:hc_real}}
   \hfill
    \subfloat[\sffamily End-to-End Performance for Cold Cache]{\includegraphics[width=0.48\linewidth]{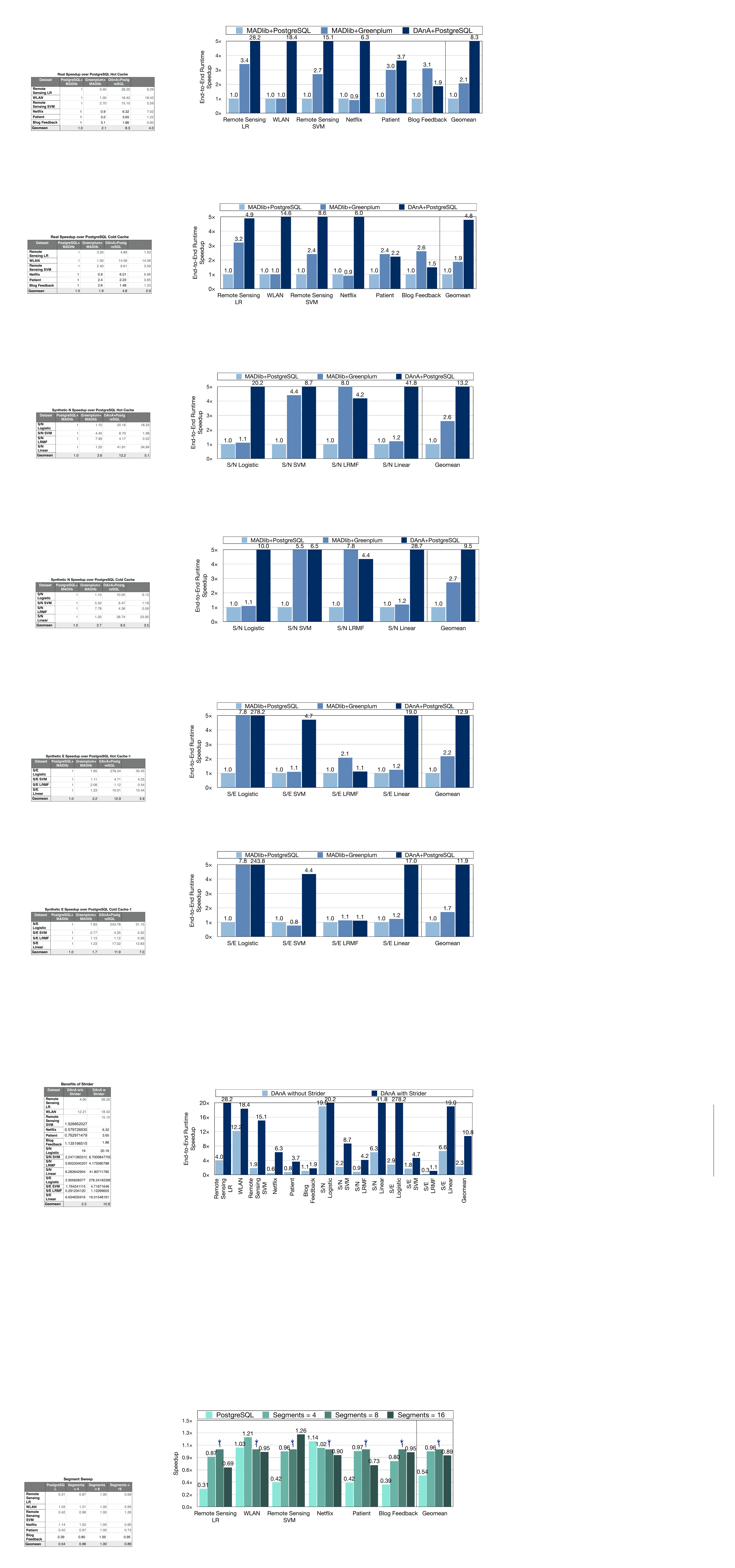}\label{fig:cc_real}}
    \vspace{-2.5ex}
  \caption{\sffamily End-to-end runtime performance comparison for publicly available datasets with \mpsql as baseline.}
  \label{fig:real}
	 \vspace{-1ex}
  %\vspace*{\floatsep}% https://tex.stackexchange.com/q/26521/5764

    \captionsetup[subfloat]{captionskip=1.6pt}
   \subfloat[\sffamily End-to-End Performance for Warm Cache]{\includegraphics[width=0.48\linewidth]{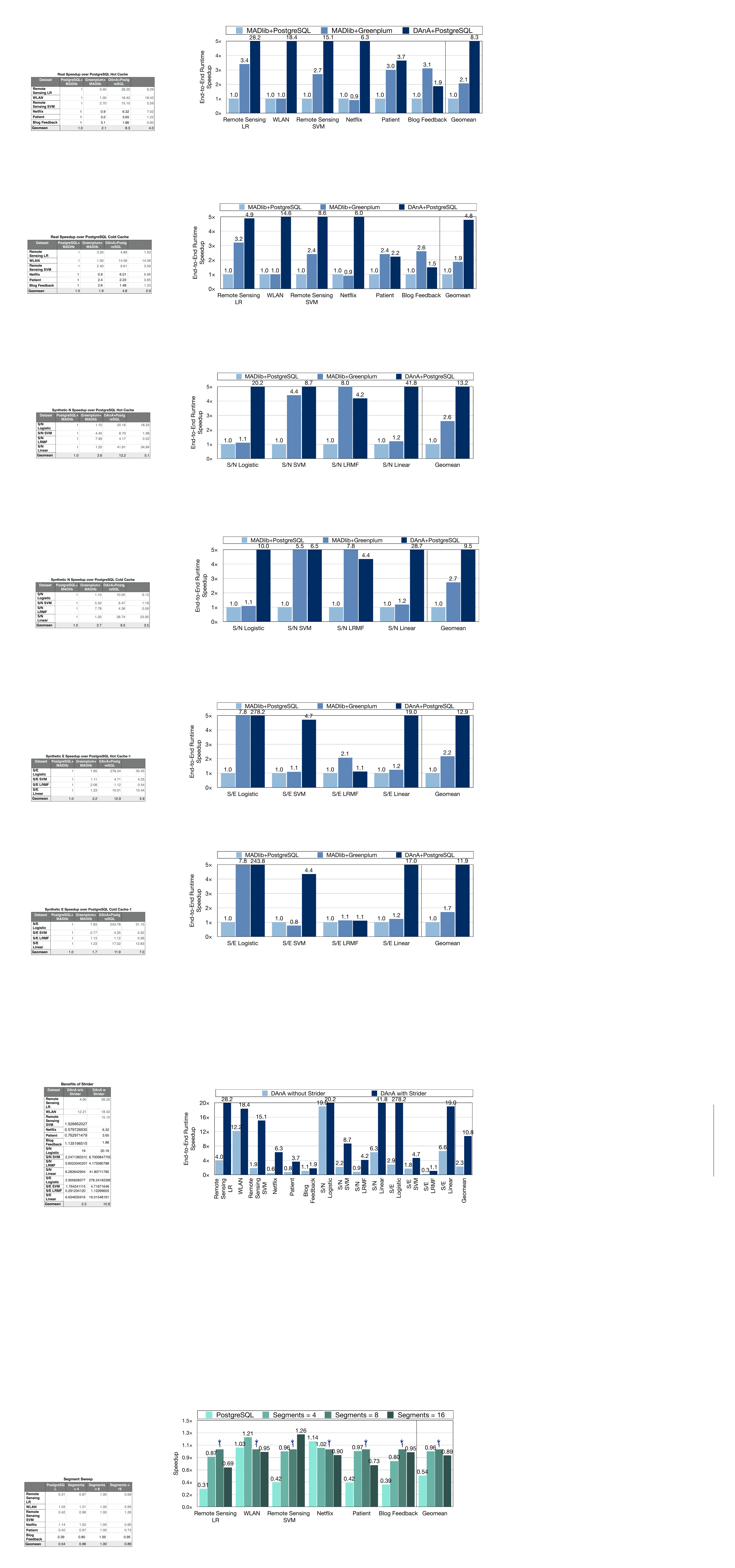} 
   \label{fig:hc_synn}}
    \hfill
    \subfloat[\sffamily End-to-End Performance for Cold Cache]{\includegraphics[width=0.48\linewidth]{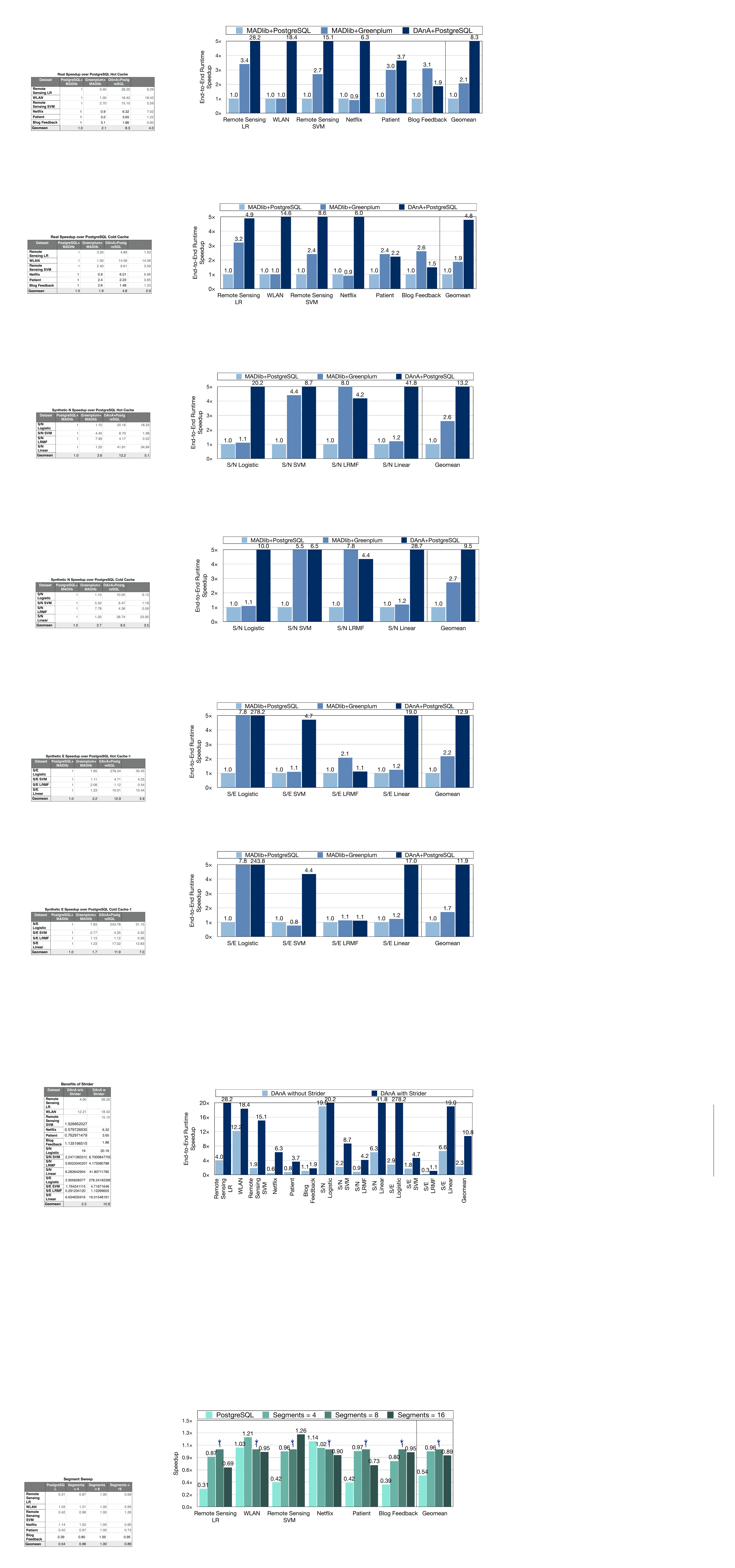} 
    \label{fig:cc_synn}}
    \vspace{-2.5ex}
  \caption{\sffamily End-to-end runtime performance comparison for synthetic nominal datasets with \mpsql as baseline.}
   \label{fig:synn}
   \vspace{-1.5ex}
   
   \captionsetup[subfloat]{captionskip=1.6pt}
   \subfloat[\sffamily End-to-End Performance for Warm Cache]{\includegraphics[width=0.48\linewidth]{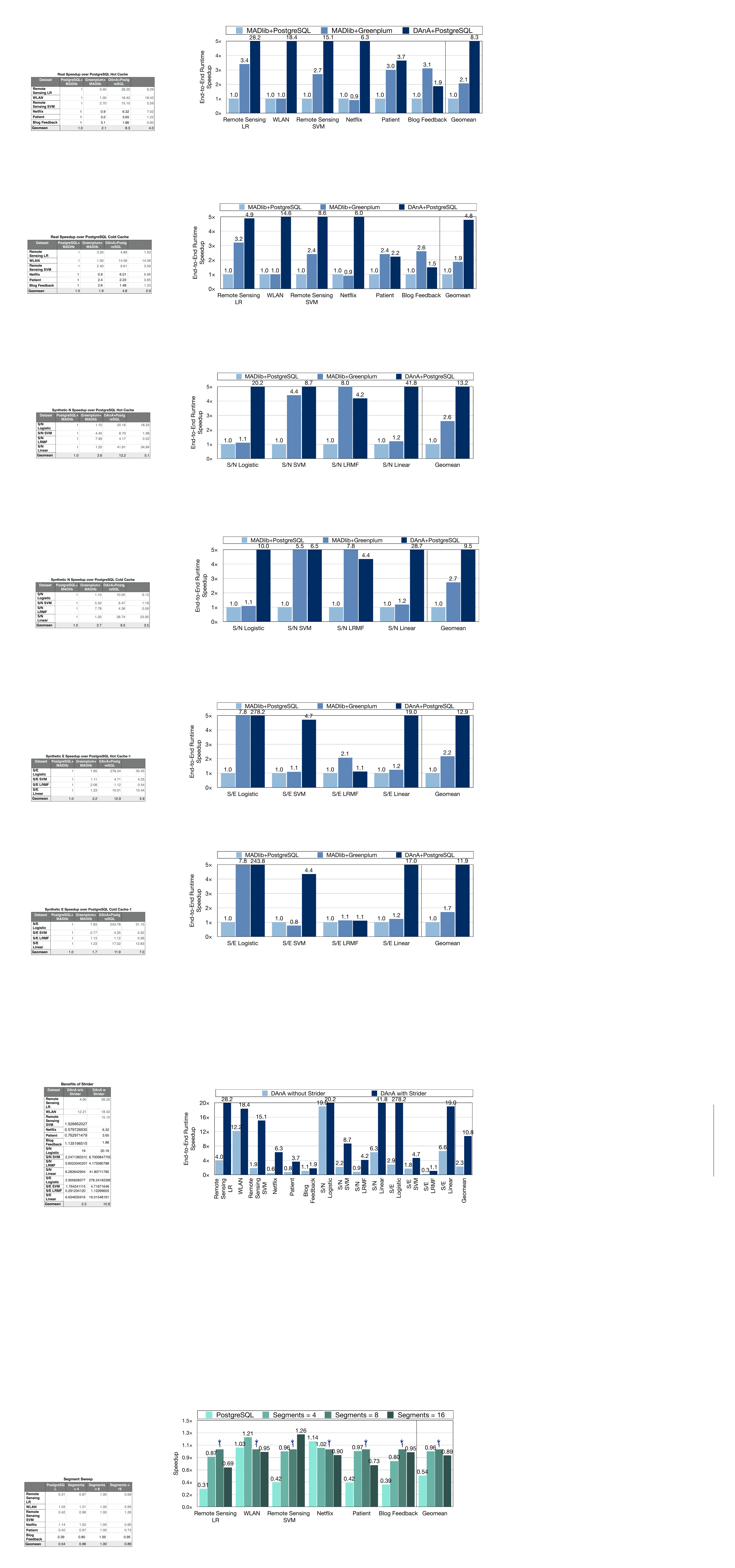} 
   \label{fig:hc_syne}}
   \hfill
    \subfloat[\sffamily End-to-End Performance for Cold Cache]{\includegraphics[width=0.48\linewidth]{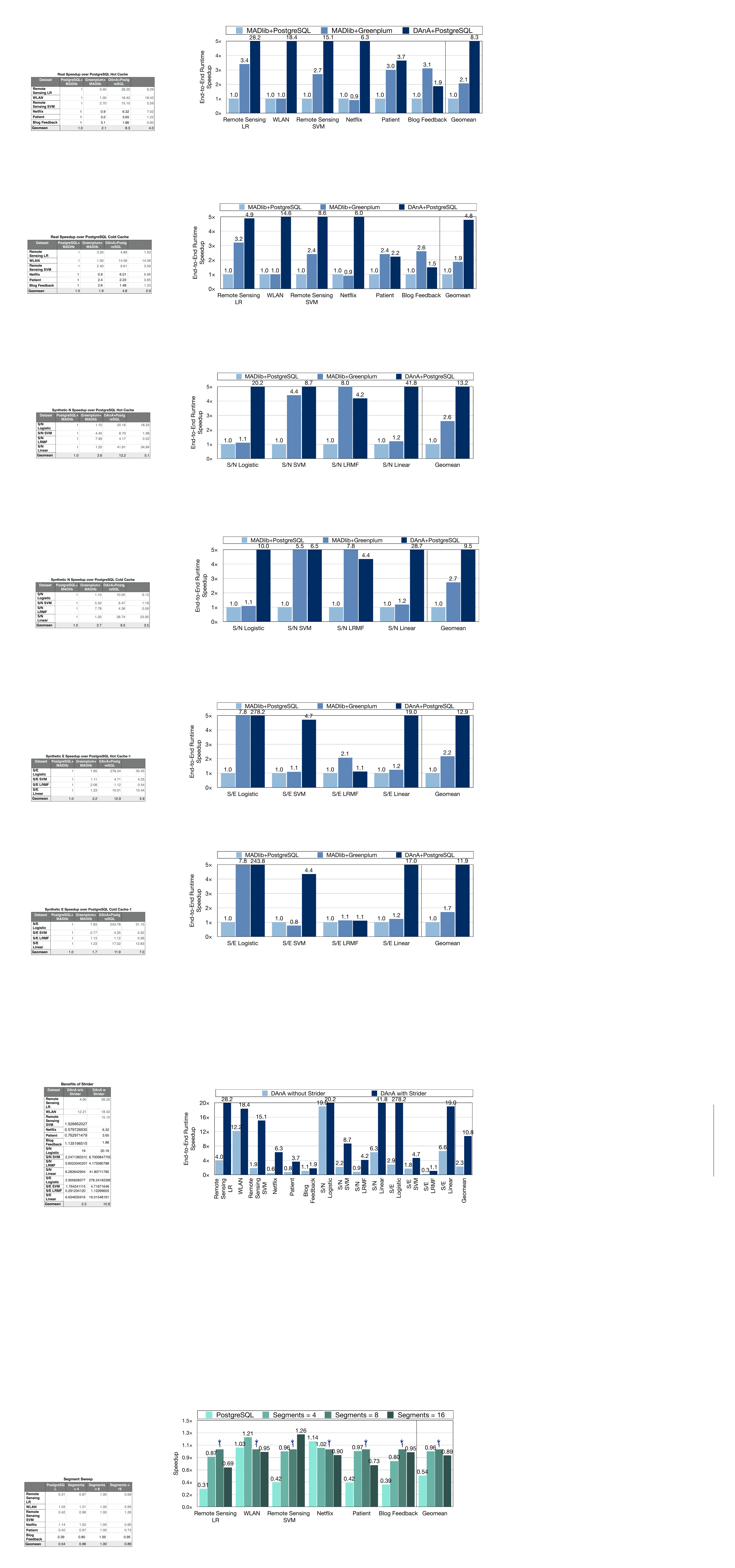} 
    \label{fig:cc_syne}}
    \vspace{-2ex}
  \caption{\sffamily End-to-end runtime performance comparison for synthetic extensive datasets with \mpsql as baseline.}
    \vspace{-3.5ex}
  \label{fig:syne}
\end{figure*}

\niparagraph{Publicly available datasets.} 
Figures~\ref{fig:hc_real} and \ref{fig:cc_real} illustrate end-to-end performance of \mpsql, Greenplum+MADlib, and \dana, for warm and cold cache. 
The x-axis represents the individual workloads and y-axis the speedup. 
The last bar provides the geometric mean (\bench{geomean}) across all workloads. 
On average, \dana provides \avgspeeduphcreal and \avgspeedupccreal end-to-end speedup over \psql and \greenavgspeeduphcreal and \greenavgspeedupccreal speedup over 8-segment Greenplum for publicly available datasets in warm and cold cache setting, respectively. 
The benefits diminish for cold cache as the I/O time adds to the runtime and cannot be parallelized. 
The overall runtime of the benchmarks reduces from 14 to 1.3 seconds with \dana in contrast to \mpsql.

The maximum speedup is obtained by \bench{Remote Sensing LR}, \maxspeeduphcreal with warm cache and \maxspeedupccreal with cold cache. 
This workload runs logistic regression algorithm to perform non-linear transformations to categorize data in different classes and offers copious amounts of parallelism for exploitation by \dana's accelerator. 
In contrast, \bench{Blog Feedback} sees the smallest speedup of~\minspeeduphcreal (warm cache) and~\minspeedupccreal (cold cache) due to the high CPU vectorization potential of the linear regression algorithm. 
%

%These workloads have a relatively large number of features but relatively fewer training data tuples. 
%
%This makes execution engine time disproportionately higher than the access engine time. 
%
%Thus, the benefits of pipelining between both engines diminishes, and execution becomes the bottleneck. 
% hence the lower speedup. 
%For Greenplum, we use the runtime measurements from the 8-segment.
%
%Using 32K pages, we can store individual tuples of all the real-world datasets in in one page and hence prevent records from spanning multiple pages, which in turn simplifies the striders' design. 

%
\niparagraph{Synthetic nominal and extensive datasets.}
Figures~\ref{fig:synn} and ~\ref{fig:syne} depict end-to-end performance comparison for synthetic nominal and extensive datasets across our three systems. 
Across \sn datasets, shown in Figure~\ref{fig:synn}, \dana achieves an average speedup of \avgspeeduphcsynn in warm cache and \avgspeedupccsynn in cold cache setting. 
In comparison to 8-segment Greenplum, for \sn datasets, \dana observes a gain of \greenavgspeeduphcsynn for warm cache and \greenavgspeedupccsynn for cold cache.
The average speedup as shown in Figure~\ref{fig:syne}, across \se datasets in comparison to \mpsql are \avgspeeduphcsyne for warm cache and \avgspeedupccsyne for cold cache.
These speedups reduce to \greenavgspeeduphcsyne (warm cache) and \greenavgspeedupccsyne (cold cache) when compared against 8-segment MADlib+Greenplum.
Higher benefits of acceleration are observed with larger datasets as \dana accelerators are exposed to more opportunities for parallelization, which enables the accelerator to hide the overheads such as data transfer across platforms, on-chip data alignment, and setting up the execution pipeline.
These results show the efficacy of the multi-threading employed by \dana's execution engine in comparison to the scale-out Greenplum engine.
The total runtime for \sn~\bench{Logistic} and \se~\bench{Logistic} reduces from 55 minutes to 2 minutes and 66 hrs to 12 minutes, respectively.
These workloads achieve high reduction in total runtimes due to their high skew towards compute time, which \dana's execution engine is specialized in handling.
%
%Runtime for \sn~\bench{LRMF} reduces from 29 minutes to 6 minutes and reduces for \se~\bench{LRMF} from 55 minutes to 39 minutes.
%
For \sn~\bench{SVM}, \dana is only able to reduce the runtime by 20 seconds.
This can be attributed to the high I/O time incurred by the benchmark in comparison to its compute time, thus, the accelerator frequently stalls for the buffer pool page replacements to complete. 
Nevertheless, for \se~\bench{SVM}, \dana still reduces the absolute runtime from 55 to 39 minutes. %
%
% I/O times, hence the execution engines can extract performance benefits.
%
%The diminishing effect of IO on speedup can be seen even more clearly in the contrast between speedup of warm and cold cache settings for the experiments on the synthetic datasets (i.e. out-of-memory workloads). In this case going from cold to warm cache does not make much difference as the RDBMS needs to IO constantly and thus the cached pages are constantly evicted and replaced. 

\begin{figure}
  \centering
   \includegraphics[width=1\linewidth]{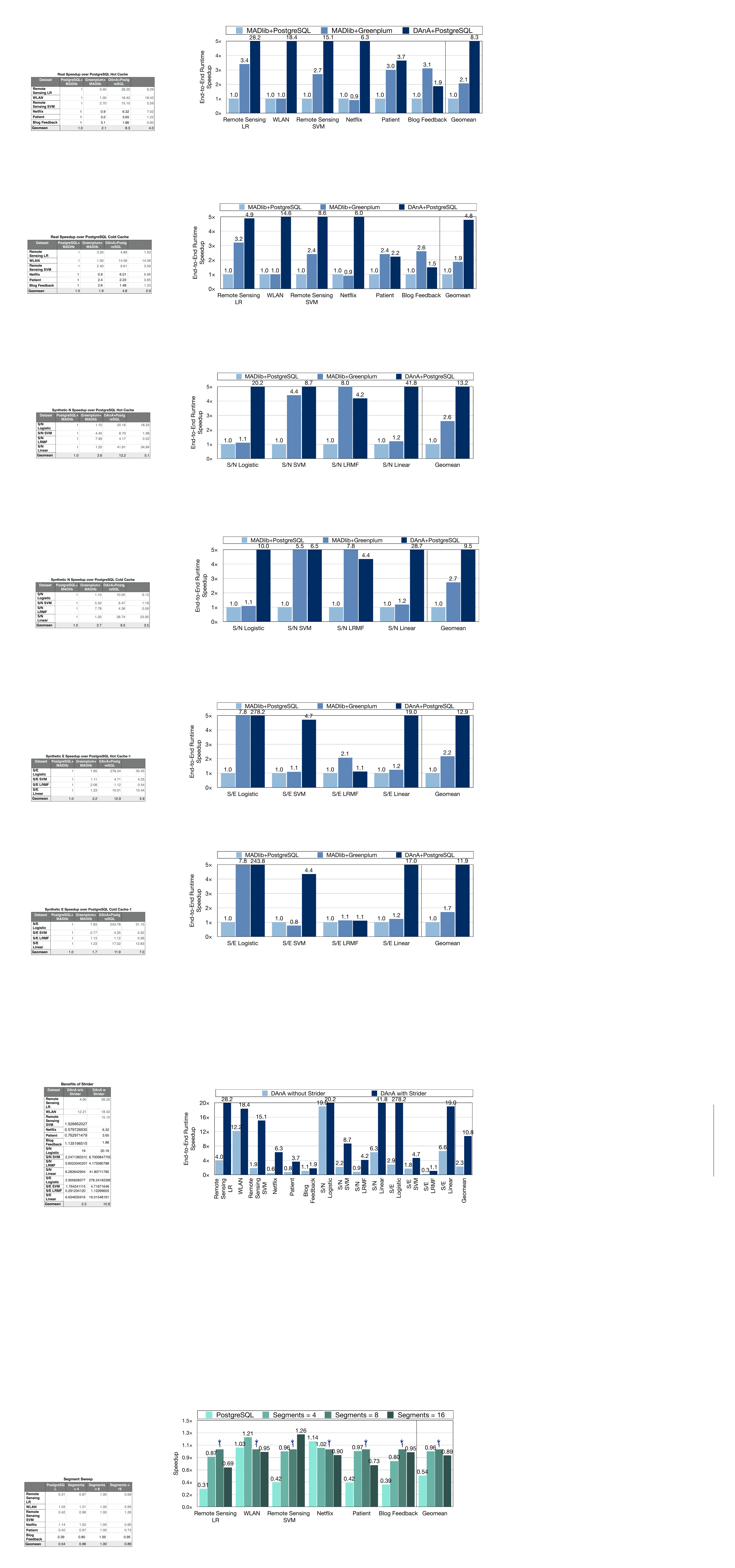} 
   \vspace{-4.5ex}
  \caption{Comparison of \dana with and without \striders with \psql+MADlib as the baseline.\label{fig:strider_benefits}}
  \vspace{-4.5ex}
\end{figure}

\begin{figure*}
  \centering
   \includegraphics[width=0.9\linewidth]{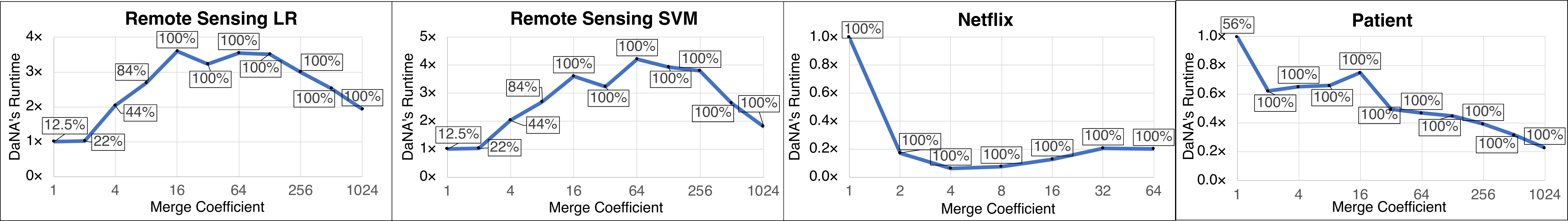} 
   \vspace{-1.5ex}
    \caption{Runtime performance of \dana with increasing \# of threads compared with single-thread as baseline.}
    \label{fig:threads}
    \vspace{-3ex}
\end{figure*}

\begin{figure}
  \centering
    \includegraphics[width=1\linewidth]{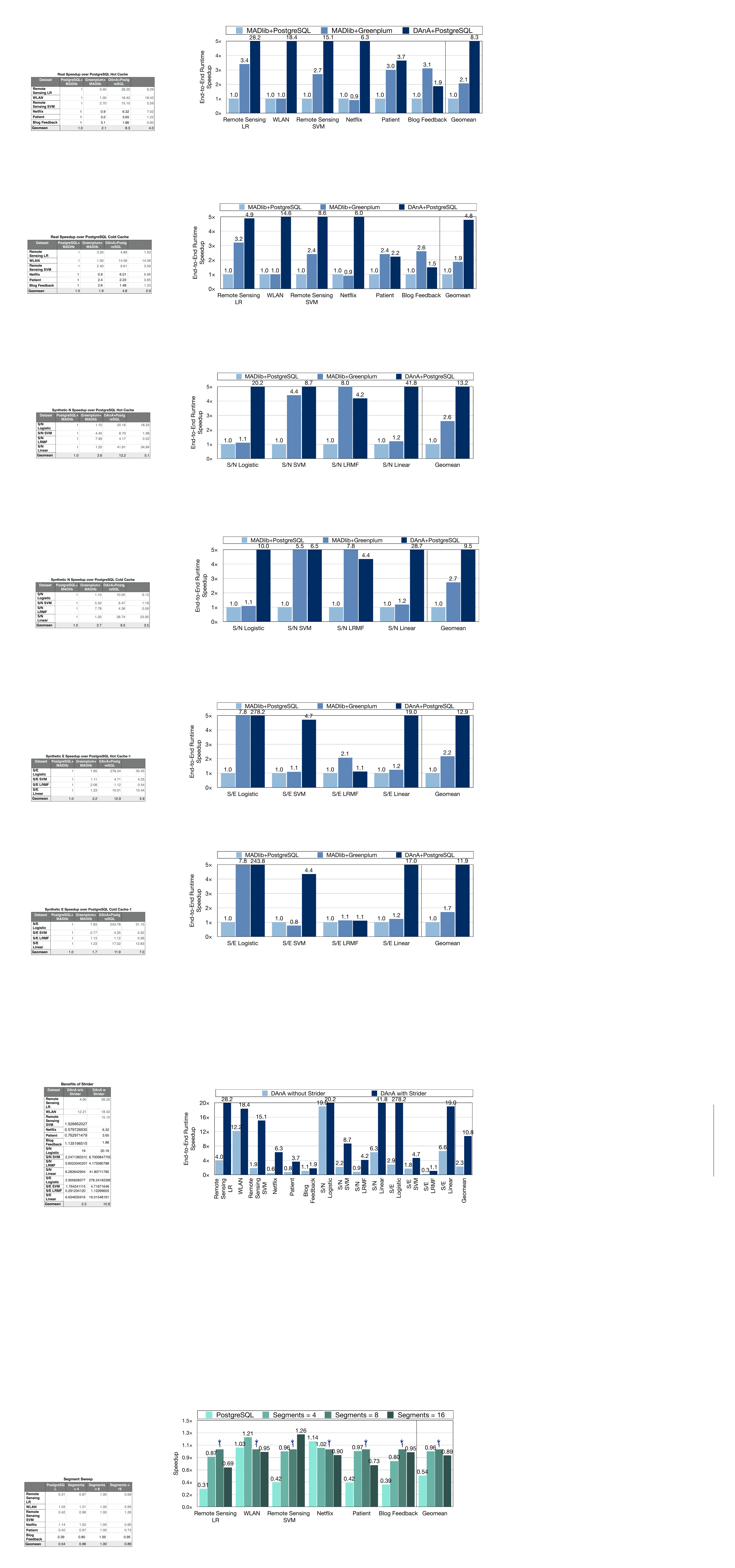}
    \vspace{-4.5ex}
  \caption{Greenplum performance with varying segments.\label{fig:segments}}
  \vspace{-2ex}
\end{figure}

\begin{figure}
  \centering
   \includegraphics[width=1\linewidth]{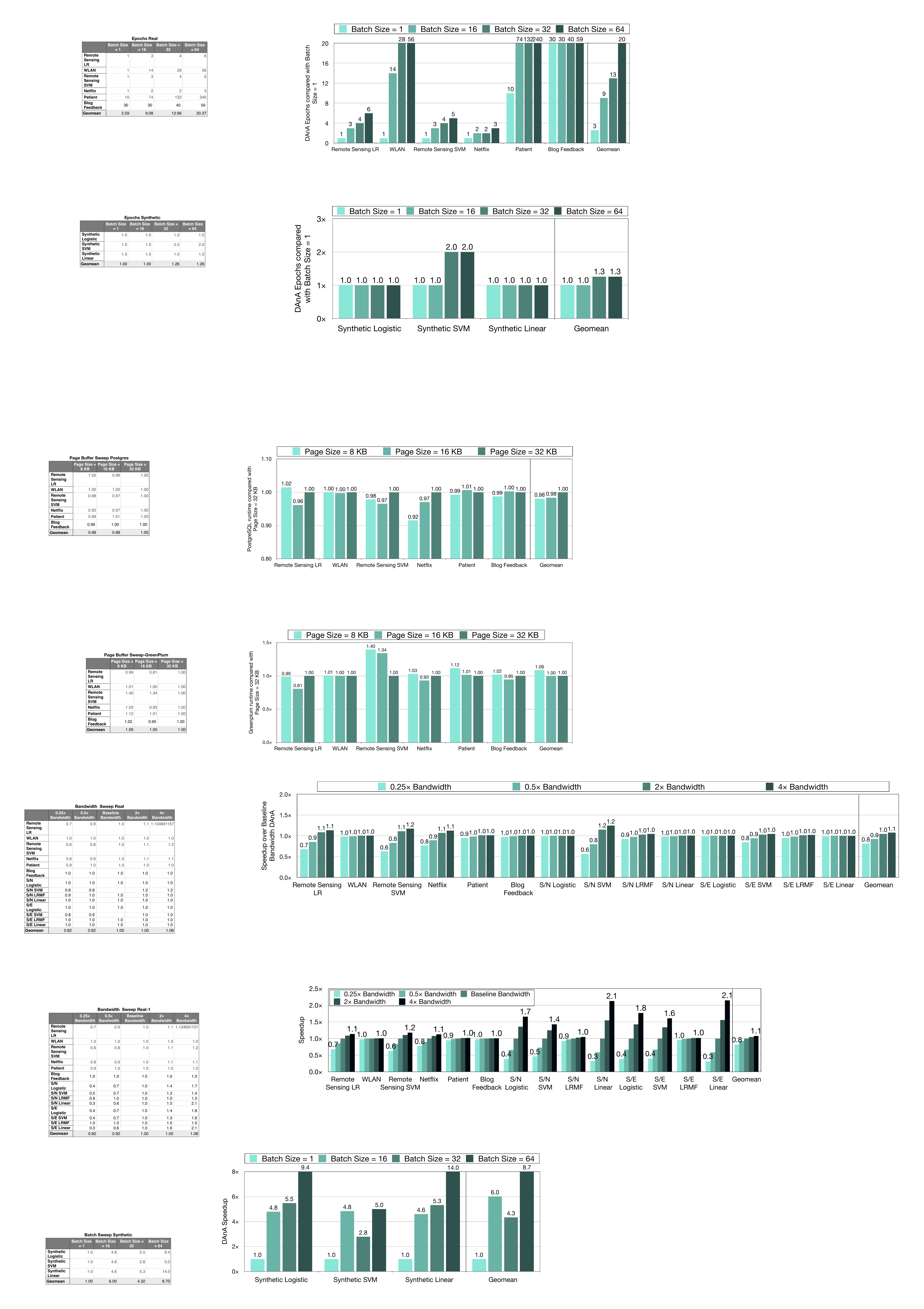} 
   \vspace{-4ex}
  \caption{Comparison of FPGA time with varying bandwidth.}
  \vspace{-4.5ex}
  \label{fig:bandwidth}
\end{figure}

\niparagraph{Evaluating \striders.}
A crucial part of \dana's accelerators is their direct integration with the buffer pool via \striders. 
To evaluate the effectiveness of \striders, we simulate an alternate design where \dana's execution engines are fed by the CPU.
In this alternative, the CPU transforms the training tuples and sends them to the execution engines.
Figure~\ref{fig:strider_benefits} compares the end-to-end runtime of \dana with and without \striders using warm cache \mpsql as baseline. 
\dana with and without \striders achieve, on average, \bench{10.7$\times$} and \avgspeeduphcrealnstrider speedup in comparison to the baseline.
Even though raw application hardware acceleration has its benefits, integrating \striders to directly interface with the database engine amplifies those performance benefits by~\danastridernstrider. 
The \striders bypass the bottlenecks in the memory subsystem of CPUs and provide an on-chip opportunity to intersperse the tasks of the access and execution engines.     
The above evaluation demonstrates the effectiveness of \dana and \striders in integrating with \psql.

%\subsection{Sensitivity Studies}
%
%This section aims to understand the impact system parameters, such as multi-threading in Greenplum and the accelerator and bandwidth of the FPGA, have on the performance of MADlib and \dana. 
%

\subsection{Performance Sensitivity}

\niparagraph{Multi-threading in Greenplum.}
As shown in Figure~\ref{fig:segments}, for publicly available datasets, the default configuration of 8-segment Greenplum provides $2.1\times$ (warm cache) and $1.9\times$ (cold cache) higher speedups than its \psql counterpart.
The 8-segment Greenplum performs the best amongst all options and performance does not scale as the segments increase. % does not significantly improve Greenplum performance. 

\niparagraph{Performance Sensitivity to FPGA resources.}
Two main resource constraints on the FPGA are its compute capability and bandwidth.
\dana configures the template architecture in accordance to the algorithmic parameters and FPGA constraints. 
To maximize compute resource utilization, \dana supports a multi-threaded execution engine, where each thread runs a version of the update rule. 
We perform an analysis for varying number of threads on the final accelerator by changing the merge coefficient.
A merge coefficient of 2 implies a maximum of two threads.
However, a large merge coefficient, such as 2048, does not warrant 2048 threads, as the FPGA may not have enough resources. 
In UltraScale+ FPGA, maximum 1024 compute units can be instantiated.
%
%The hardware generator strikes a balance between the resources allocated to a single thread vs. number of threads to obtain the maximum throughput. 

Figure~\ref{fig:threads} shows performance sensitivity with increasing  compute utilization of FPGA for different workloads.
Each plot shows \dana's accelerator runtime (access engine + execution engine) in comparison to a single-thread. 
The sensitivity towards compute resources is a function of algorithm type, model width, and \# of epochs.
Thus, each workload fares differently with varying compute resources. 
Workloads such as~\bench{Remote Sensing LR} and ~\bench{Remote Sensing SVM} have a narrow model size, thus, increasing the number of threads increases performance till they reach peak compute utilization.
On the other hand, LRMF algorithm workloads do not experience a higher performance  with increasing number of threads.
This can be attributed to the copious amounts of parallelism available in a single instance of the update rule. 
Thus, increasing the number of threads reduces the ACs allocated to a single thread, whereas, merging across multiple different threads incurs an overhead.
One of the challenges tackled by the compiler is to allocate the on-chip resources by striking a balance between the single-thread performance and multi-thread parallelism.

Figure~\ref{fig:bandwidth} illustrates the impact of FPGA bandwidth (in comparison to baseline bandwidth) on the speedup of \dana accelerators.
The results show that as the size of the benchmark increases, except the ones that run LRMF algorithm, the workloads become bandwidth bound.
The workloads \sn~\bench{LRMF} and \se~\bench{LRMF} are compute heavy, thus, bandwidth increase does not have a significant impact on the accelerator runtime.

\subsection{Comparison to Custom Designs}

Potential alternatives to \dana are: (1) custom software libraries~\cite{strads, liblinear, dimmwitted} that run multi-core advanced analytics and (2) algorithm specific FPGA implementations~\cite{svm1, svm2, lgr1}. 
For these alternatives, if training data is stored in the database, there is an overhead to extract, transform, and supply the data in accordance to each of their requirements.
We compare the performance of these alternatives with \mpsql and \dana.

\begin{figure}
\captionsetup[subfloat]{captionskip=3pt}
  \centering
      \subfloat[\sffamily Runtime breakdown for Liblinear and DimmWitted]{\includegraphics[width=1\linewidth]{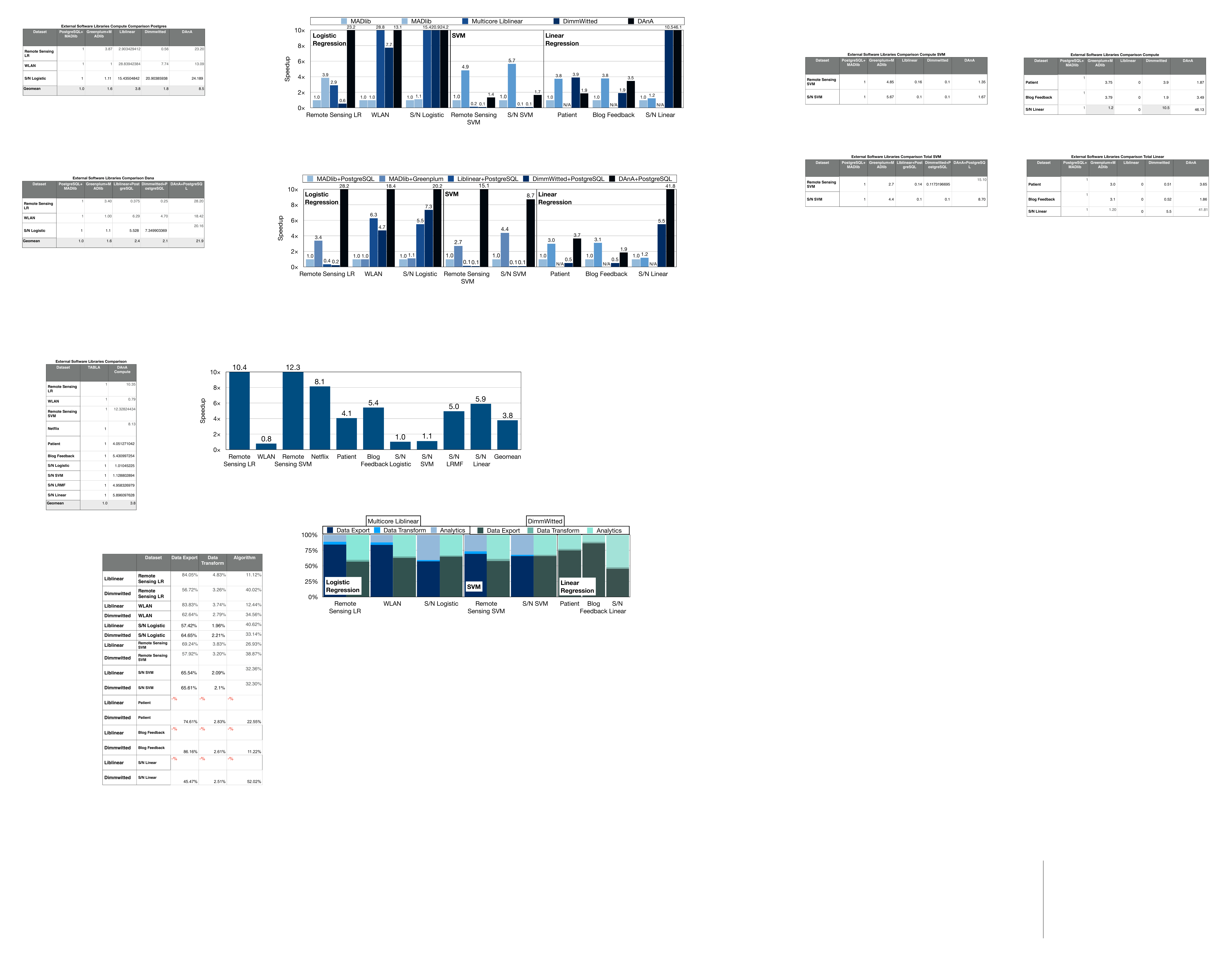} \label{fig:externallibsbreakdown}}
  \vspace{-1.5ex}
  \newline
    \subfloat[\sffamily Compute Time Comparison]{\includegraphics[width=1\linewidth]{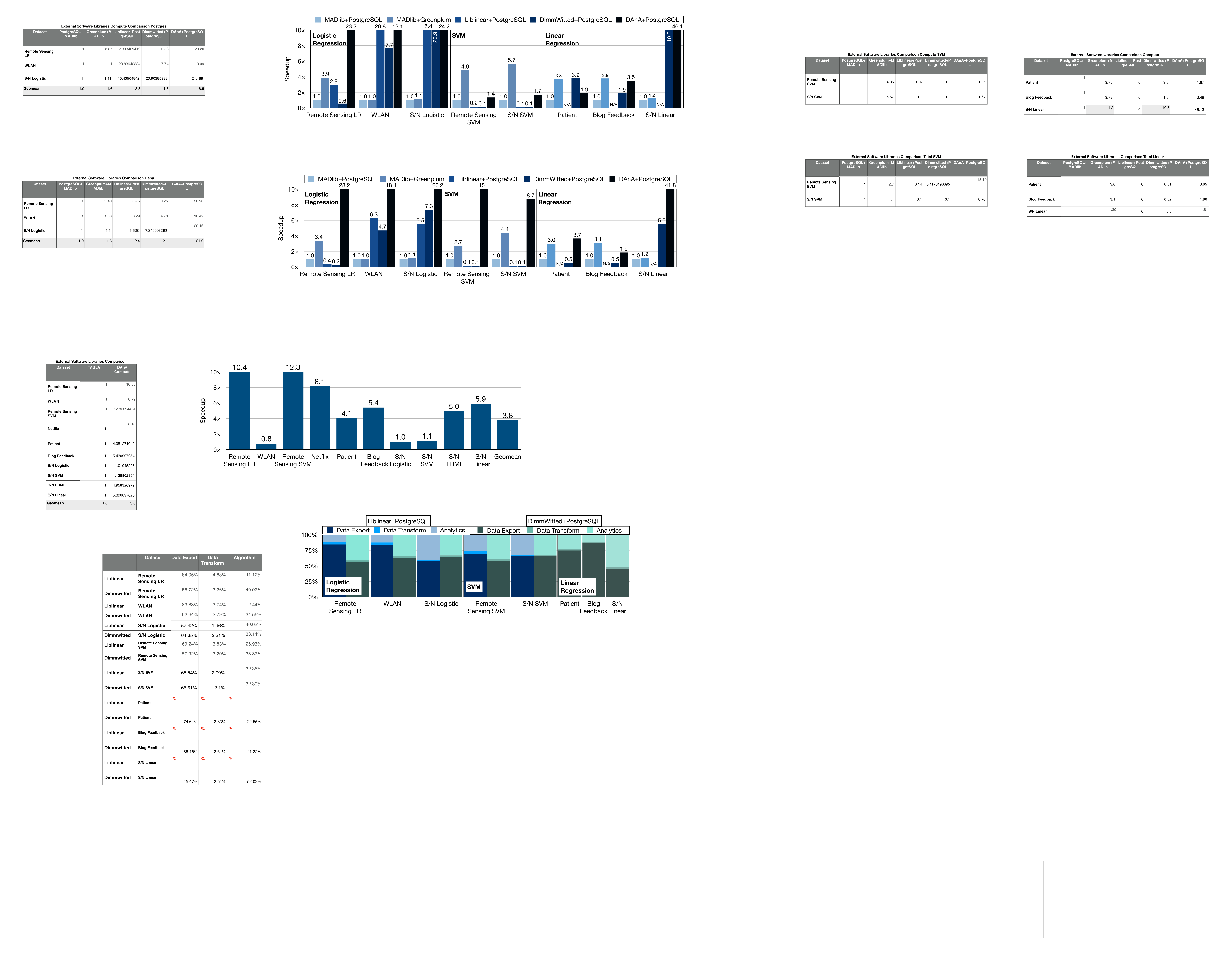} \label{fig:externallibscompute}}
    \vspace{-1.5ex}
    \newline
    \subfloat[\sffamily End-to-End Runtime Comparison]{\includegraphics[width=1\linewidth]{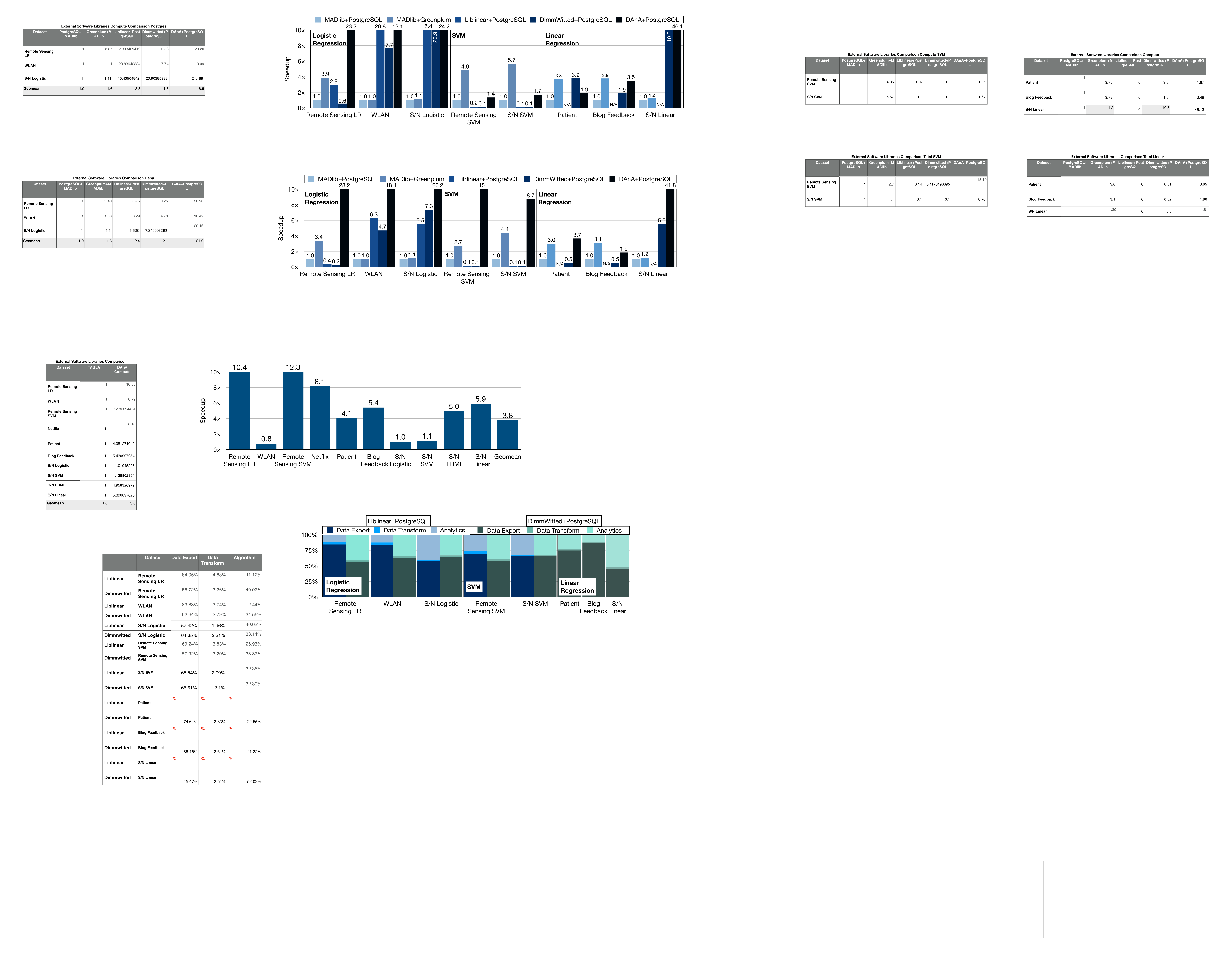} \label{fig:externallibstotal}}
   \vspace{-1.7ex}
  \caption{Comparison to external software libraries. \label{fig:externallibs}}
  \vspace{-4.5ex}
\end{figure} 

\niparagraph{Optimized software libraries.}
We compare C++-optimized libraries DimmWitted and Liblinear-Multicore classification with \mpsql,  Greenplum+MADlib, and \dana accelerators. 
Liblinear supports Logistic Regression and SVM, and DimmWitted supports SVM, Logistic Regression, Linear Regression, Linear Programming, Quadratic Programming, Gibbs Sampling, and Neural Networks.
Logistic Regression, SVM, and Linear Regression (only DimmWitted), overlap with our benchmarks, thus, we compare multi-core versions (2, 4, 8, 16 threads) of these libraries and use the minimum runtime.
We maintain the same hyper-parameters, such as tolerance, and choice of optimizer to compare runtime of 1 epoch across all the systems.
We separately compare the compute time and end-to-end runtime (data extraction from \psql + data transformation + compute), as well as provide a runtime breakdown.
Figure~\ref{fig:externallibsbreakdown} illustrates the breakdown of Liblinear and DimmWitted into the different phases that comprise the end-to-end runtime.
Data exporting and reformatting for these external specialized ML tools is an overhead specific to performing analytics outside RDBMS. 
Results suggest that \dana is uniformly faster, as it (1) does not export the data from the database, (2) employs \striders in the FPGA to walk through the data, and (3) accelerates the ML computation with an FPGA.
However, different software solutions exhibit different trends, as elaborated below.

\begin{itempacked}

\item \textbf{Logistic regression:} As Figure \ref{fig:externallibscompute} shows, Liblinear and DimmWitted provide~\bench{3.8$\times$} and~\bench{1.8$\times$} speedup over \mpsql for logistic regression-based workloads in terms of compute time.
With respect to end-to-end runtime compared to \mpsql (Figure~\ref{fig:externallibstotal}), the benefits from Liblinear reduce to \bench{2.4$\times$} and increase for DimmWitted to \bench{2.1$\times$}.
On the other hand, \dana outperforms Liblinear by \bench{2.2$\times$} and DimmWitted by \bench{4.7$\times$} in terms of compute time. 
For overall runtime, \dana is \bench{9.1$\times$} faster than Liblinear and \bench{10.4$\times$} faster than DimmWitted.
The compute time of \bench{Remote sensing LR} benchmark receives the least benefit from LibLinear and DimmWitted and exhibits a slowdown in end-to-end runtime. 
This can be attributed to the small model size, which, despite a large dataset, does not provide enough parallelism that can be exploited by these libraries.
Specifically sparse datasets, such as \bench{WLAN}, are handled more efficiently by these libraries.

\item \textbf{SVM:} 
As shown in Figure \ref{fig:externallibscompute}, that compares compute time of SVM-based workloads, Liblinear and DimmWitted are~\bench{18.1$\times$} and ~\bench{22.3$\times$} slower than \mpsql, respectively.
For end-to-end runtime (Figure~\ref{fig:externallibstotal}), the slowdown is reduced to \bench{14.6$\times$} for Liblinear  and \bench{15.9$\times$} for DimmWitted, due to the complex interplay between data accesses and UDF execution of \mpsql.
In comparison, \dana outperforms Liblinear by \bench{30.7$\times$} and DimmWitted by \bench{37.7$\times$} in terms of compute time.
For overall runtime, \dana is \bench{127$\times$} and \bench{138.3$\times$} faster than Liblinear and DimmWitted, respectively.

\item \textbf{Linear regression:}
For linear regression-based workloads, DimmWitted is~\bench{4.3$\times$} faster than \mpsql in terms of compute time.
For end-to-end runtime compared to \mpsql (Figure~\ref{fig:externallibstotal}), the speedup of DimmWitted is reduced to \bench{12\%}.
\dana outperforms DimmWitted by \bench{1.6$\times$} and \bench{6.0$\times$} in terms of compute and overall time, respectively.

\end{itempacked}

\niparagraph{Specific FPGA implementations.} %and High Level Synthesis.}}
We compare hand-optimized FPGA designs created specifically for one algorithm with our reconfigurable architecture. 
%
%These custom designs provide support for SVM and logistic regression algorithms.
%
\dana's execution engine performance is on par with Parallel SVM~\cite{svm1}, is ~\bench{44\%} slower than Heterogeneous SVM~\cite{svm2}, and is~\bench{1.47$\times$} faster than Falcon Logistic Regression~\cite{lgr1}.
In addition to the speedup, we compare Giga Ops Per Second (GOPS), to measure the numerical compute performance of these architectures. 
In terms of GOPS, \dana performs, on average, 16\% less operations than these hand-coded designs.
In addition to providing comparable performance, \dana relieves the data scientist of the arduous task of hardware design and testing whilst integrating seamlessly within the database engine. 
Whereas, for these custom designs, designer requires hardware design expertise and long verification cycles to write $\approx$15000 lines of Verilog code.

\begin{figure}
  \centering
   \includegraphics[width=1\linewidth]{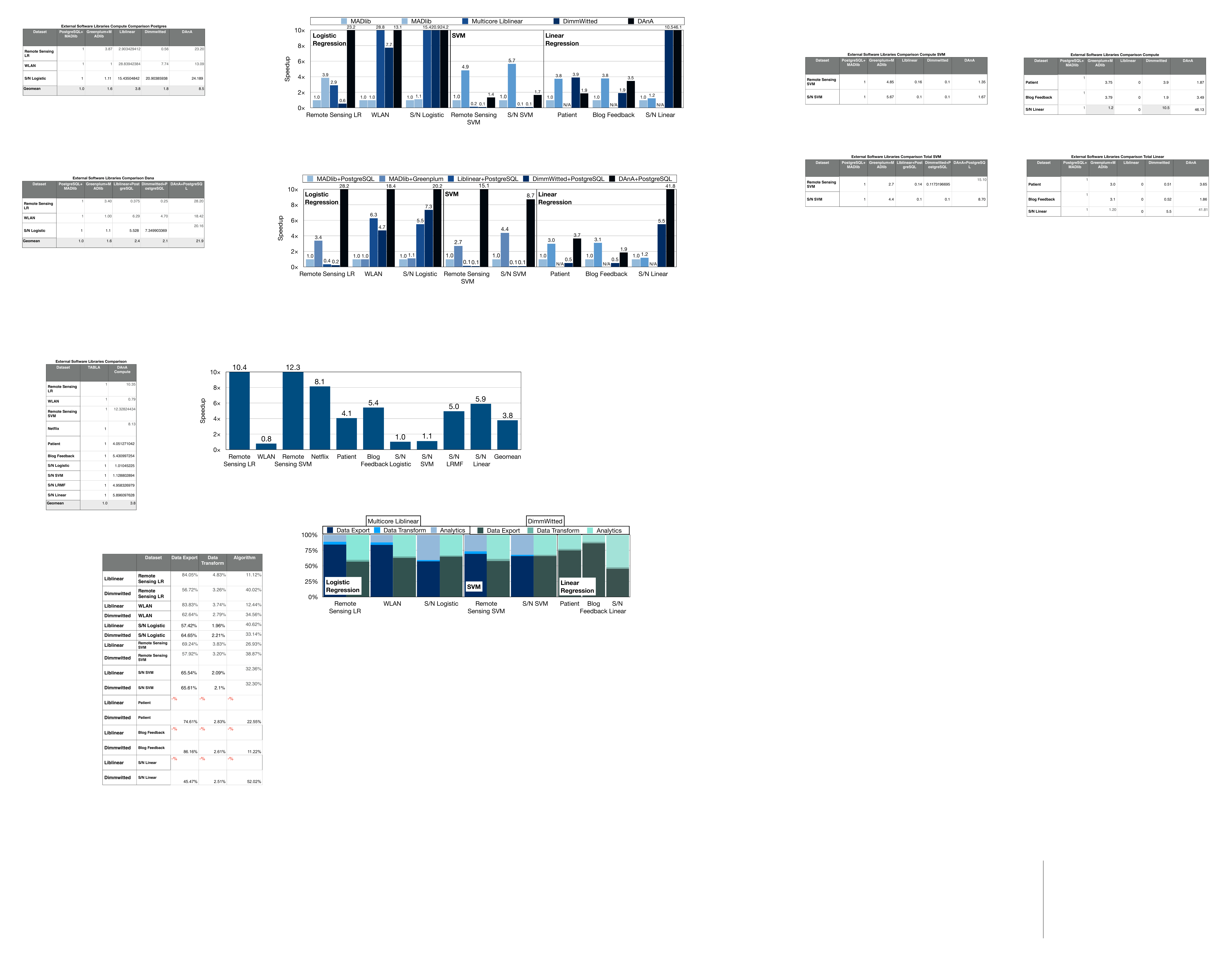} 
   \vspace{-4ex}
  \caption{Performance comparison of \dana over \captabla.\label{fig:tabla}}
  \vspace{-4.5ex}
  \label{fig:tabla}
\end{figure}

\niparagraph{Comparison with \captabla.}
We compare \dana with \tabla~\cite{tabla:hpca}, an open-source framework~\cite{tabla-opensource} that generates optimized FPGA implementations for a wide variety of analytics algorithms.
We modify the templates for UltraScale+ and perform design space exploration to present the best case results with \tabla.
Figure~\ref{fig:tabla} shows that \dana generated accelerators perform \tablaspeedup faster than \tabla accelerators. 
\dana's benefits can be attributed to the: (1) interleaving of \striders in the access engine with the execution engine to mitigate the overheads of data transformation and (2) the multi-threading capability of the execution engines to exploit parallelism between different instances of the update rule.  
\tabla on the other hand, offers only single threaded acceleration.
%
%For the workload ~\bench{WLAN}, \tabla outperforms \dana because it is sparse dataset, hence increasing the number of threads exponentially increases the epochs required for converging. 

\section{Related Work}

\niparagraph{Hardware acceleration for data management.}
Accelerating database operations is a popular research direction that connects modern acceleration platforms and enterprise in-database analytics as shown in Figure~\ref{fig:triad}.
%
%In contrast, we focus on accelerating in-RDBMS advanced analytics through ML.
%
These prior FPGA-based solutions aim to accelerate DBMS operations (some portion of the query)~\cite{centaur, doppiodb, gustavofpga, fpgapartitioning, linqits}, such as join and hash.
%
%Generally, these works focus on utilizing FPGA for some portion of the query flow while the rest runs in the CPU.
%
LINQits~\cite{linqits} accelerates database queries but does not focus on machine learning.
Centaur \cite{centaur} dynamically decides which particular operators in a MonetDB \cite{monetdb} query plan can be executed on FPGA and creates a pipeline between FPGA and CPU.
Another work~\cite{fpgapartitioning} uses FPGAs to provide a robust hashing mechanism to accelerate data partitioning in database engines. 
In the GPU realm, HippogriffDB~\cite{hippogriffdb} aims to balance the I/O and GPU bandwidth by compressing the data that is transferred to GPU.
%
%These work does not deal with advanced analytics. %, although GPUs are an appropriate platform.
%
Support for in-database advanced analytics for FPGAs in tandem with \striders set this work apart from the aforementioned literature, which does not focus on providing components that integrate FPGAs within an RDBMS engine and machine learning.  
%
%\dana on the other hand, deeply integrates with the RDBMS engine by bypassing the CPU, with a focus on in-Database advanced analytics.

%
 \niparagraph{Hardware acceleration for advanced analytics.}
Both research and industry have recently focused on hardware acceleration for machine learning~\cite{pudiannao, tabla:hpca, diannao, gustavoce} especially deep neural networks~\cite{shidiannao, eie:isca2016, eyeriss, cnvlutin, minerva:isca2016} connecting two of the vertices in Figure~\ref{fig:triad} traid.
These works either only focus on a fixed set of algorithms or do not offer the reconfigurability of the architecture.
Among these, several works~\cite{tabla, dnnweaver, cosmic:micro} provide frameworks to automatically generate hardware accelerators for stochastic gradient descent.
However, none of these works provide hardware structures or software components that embed FPGAs within the RDBMS engine.  
\dana's Python DSL builds upon the mathematical language in the prior work~\cite{tabla:hpca, cosmic:micro}.
However, the integration with both conventional (Python) and data access (SQL) languages provides a significant extension by enabling support for UDFs which include general iterative update rules, merge functions, and convergence functions. 

\niparagraph{In-Database advanced analytics.}
Recent work at the intersection of databases and machine learning are extensively trying to facilitate efficient in-database analytics and have built frameworks and systems to realize such an integration \cite{oraclesvm, crfindb, mcmcindb, bismarck, madlib1, madlib2, shark, hazyclassification, hogwild, tupleware, weld} (see \cite{dmml} for a survey of various methods and systems). 
\dana takes a step forward and exposes FPGA acceleration for in-Database analytics by providing a specialized component, \strider, that directly interfaces with the database to alleviate some of the shortcomings of the traditional Von-Neumann architecture in general purpose compute systems.
Past work in Bismarck~\cite{bismarck} provides a unified architecture for in-database analytics, facilitating UDFs as an interface for the analyst to describe their desired analytics models.
However, unlike \dana, Bismarck lacks the hardware acceleration backend and support for general iterative optimization algorithms.
%
%We build on our previous work, Bismarck~\cite{bismarck}, which offers a unified architecture for in-database analytics, facilitating UDFs as a convenient interface for the analyst to describe their desired analytics models.
%

%\niparagraph{Automated Frameworks for FPGA Acceleration.}
%
%To the best of our knowledge, \dana is the first extensive work that exposes modern hardware FPGA platform for in-database analytics by providing a familiar application programing interface (Python and SQL) to the enterprise analysts.
\section{Conclusion}
\label{conclusion}
This paper aims to bridge the power of well-established and extensively researched means of structuring, defining, protecting, and accessing data, i.e., RDBMS with FPGA accelerators for compute-intensive advanced data analytics.
\dana provides the initial coalescence between these paradigms and empowers data scientists with no knowledge of hardware design, to use accelerators within their current in-database analytics procedures.
\section{Acknowledgments}
We thank Sridhar Krishnamurthy and Xilinx for donating the FPGA boards. 
We thank Yannis Papakonstantinou, Jongse Park, Hardik Sharma, and Balaji Chandrasekaran for providing insightful feedback. 
This work was in part supported by NSF awards CNS\#1703812, ECCS\#1609823, Air Force Office of Scientific Research (AFOSR) Young Investigator Program (YIP) award \#FA9550-17-1-0274, and gifts from Google, Microsoft, Xilinx, Qualcomm, and Opera Solutions.
\balance

%\end{document}  % This is where a 'short' article might terminate
% ensure same length columns on last page (might need two sub-sequent latex runs)
%\balance

\bibliographystyle{unsrt}
\bibliography{ms}{}  % vldb_sample.bib is the name of the Bibliography in this case

%\input{appendix.tex}

\end{document}